\documentclass[ALICE,manyauthors]{cernphprep}
\usepackage[comma,square,numbers,sort&compress]{natbib}

\usepackage{lineno}
\usepackage{xspace}
\usepackage{xcolor}
\usepackage{hyperref}
\usepackage{lineno}
\usepackage[T1]{fontenc}
\usepackage{orcidlink}
\begin{document}
%%%%%%%%%%%%%%%%%%%%%%%%%%%%%%%%%%%%%%%%%%%%%%%%%%
% These are some new commands that may be useful 
% for paper writing in general. If other newcommands
% are needed for your specific paper, please feel 
% free to add here. 
%
% The currently available commands are organized in: 
% 1) Systems
% 2) Quantities
% 3) Energies and units
% 4) Detectors
% 5) particle species 
%%%%%%%%%%%%%%%%%%%%%%%%%%%%%%%%%%%%%%%%%%%%%%%%%%

% 1) SYSTEMS 
\newcommand{\pp}           {pp\xspace}
\newcommand{\ppbar}        {\mbox{$\mathrm {p\overline{p}}$}\xspace}
\newcommand{\XeXe}         {\mbox{Xe--Xe}\xspace}
\newcommand{\PbPb}         {\mbox{Pb--Pb}\xspace}
\newcommand{\pA}           {\mbox{pA}\xspace}
\newcommand{\pPb}          {\mbox{p--Pb}\xspace}
\newcommand{\AuAu}         {\mbox{Au--Au}\xspace}
\newcommand{\dAu}          {\mbox{d--Au}\xspace}

% 2) QUANTITIES 
\newcommand{\s}            {\ensuremath{\sqrt{s}}\xspace}
\newcommand{\snn}          {\ensuremath{\sqrt{s_{\mathrm{NN}}}}\xspace}
\newcommand{\pt}           {\ensuremath{p_{\rm T}}\xspace}
\newcommand{\meanpt}       {$\langle p_{\mathrm{T}}\rangle$\xspace}
\newcommand{\ycms}         {\ensuremath{y_{\rm CMS}}\xspace}
\newcommand{\ylab}         {\ensuremath{y_{\rm lab}}\xspace}
\newcommand{\y}         {\ensuremath{y}\xspace}
\newcommand{\etarange}[1]  {\mbox{$\left | \eta \right |~<~#1$}}
\newcommand{\yrange}[1]    {\mbox{$\left | y \right |~<~#1$}}
\newcommand{\dndy}         {\ensuremath{\mathrm{d}N_\mathrm{ch}/\mathrm{d}y}\xspace}
\newcommand{\dndeta}       {\ensuremath{\mathrm{d}N_\mathrm{ch}/\mathrm{d}\eta}\xspace}
\newcommand{\avdndeta}     {\ensuremath{\langle\dndeta\rangle}\xspace}
\newcommand{\dNdy}         {\ensuremath{\mathrm{d}N_\mathrm{ch}/\mathrm{d}y}\xspace}
\newcommand{\Npart}        {\ensuremath{N_\mathrm{part}}\xspace}
\newcommand{\Ncoll}        {\ensuremath{N_\mathrm{coll}}\xspace}
\newcommand{\dEdx}         {\ensuremath{\textrm{d}E/\textrm{d}x}\xspace}
\newcommand{\RpPb}         {\ensuremath{R_{\rm pPb}}\xspace}

% 3) ENERGIES, UNITS
\newcommand{\nineH}        {$\sqrt{s}~=~0.9$~Te\kern-.1emV\xspace}
\newcommand{\seven}        {$\sqrt{s}~=~7$~Te\kern-.1emV\xspace}
\newcommand{\eight}        {$\sqrt{s}~=~8$~Te\kern-.1emV\xspace}
\newcommand{\twoH}         {$\sqrt{s}~=~0.2$~Te\kern-.1emV\xspace}
\newcommand{\twosevensix}  {$\sqrt{s}~=~2.76$~Te\kern-.1emV\xspace}
\newcommand{\five}         {$\sqrt{s}~=~5.02$~Te\kern-.1emV\xspace}
\newcommand{\fiveExactly}  {$\sqrt{s}~=~5$~Te\kern-.1emV\xspace}
\newcommand{\twosevensixnn}{$\sqrt{s_{\mathrm{NN}}}~=~2.76$~Te\kern-.1emV\xspace}
\newcommand{\fivenn}       {$\sqrt{s_{\mathrm{NN}}}~=~5.02$~Te\kern-.1emV\xspace}
\newcommand{\LT}           {L{\'e}vy-Tsallis\xspace}
\newcommand{\GeVc}         {Ge\kern-.1emV/$c$\xspace}
\newcommand{\MeVc}         {Me\kern-.1emV/$c$\xspace}
\newcommand{\TeV}          {Te\kern-.1emV\xspace}
\newcommand{\GeV}          {Ge\kern-.1emV\xspace}
\newcommand{\GeVtwo}       {Ge\kern-.1emV$^2$\xspace}
\newcommand{\MeV}          {Me\kern-.1emV\xspace}
\newcommand{\GeVmass}      {Ge\kern-.2emV/$c^2$\xspace}
\newcommand{\MeVmass}      {Me\kern-.2emV/$c^2$\xspace}
\newcommand{\lumi}         {\ensuremath{\mathcal{L}}\xspace}

% 4) DETECTORS 
\newcommand{\ITS}          {\rm{ITS}\xspace}
\newcommand{\TOF}          {\rm{TOF}\xspace}
\newcommand{\ZDC}          {\rm{ZDC}\xspace}
\newcommand{\ZDCs}         {\rm{ZDCs}\xspace}
\newcommand{\ZNA}          {\rm{ZNA}\xspace}
\newcommand{\ZNC}          {\rm{ZNC}\xspace}
\newcommand{\SPD}          {\rm{SPD}\xspace}
\newcommand{\SDD}          {\rm{SDD}\xspace}
\newcommand{\SSD}          {\rm{SSD}\xspace}
\newcommand{\TPC}          {\rm{TPC}\xspace}
\newcommand{\TRD}          {\rm{TRD}\xspace}
\newcommand{\VZERO}        {\rm{V0}\xspace}
\newcommand{\VZEROA}       {\rm{V0A}\xspace}
\newcommand{\VZEROC}       {\rm{V0C}\xspace}
\newcommand{\Vdecay} 	   {\ensuremath{V^{0}}\xspace}

% 4) PARTICLE SPECIES 
\newcommand{\ee}           {\ensuremath{e^{+}e^{-}}} 
\newcommand{\pip}          {\ensuremath{\pi^{+}}\xspace}
\newcommand{\pim}          {\ensuremath{\pi^{-}}\xspace}
\newcommand{\kap}          {\ensuremath{\rm{K}^{+}}\xspace}
\newcommand{\kam}          {\ensuremath{\rm{K}^{-}}\xspace}
\newcommand{\pbar}         {\ensuremath{\rm\overline{p}}\xspace}
\newcommand{\kzero}        {\ensuremath{{\rm K}^{0}_{\rm{S}}}\xspace}
\newcommand{\lmb}          {\ensuremath{\Lambda}\xspace}
\newcommand{\almb}         {\ensuremath{\overline{\Lambda}}\xspace}
\newcommand{\Om}           {\ensuremath{\Omega^-}\xspace}
\newcommand{\Mo}           {\ensuremath{\overline{\Omega}^+}\xspace}
\newcommand{\X}            {\ensuremath{\Xi^-}\xspace}
\newcommand{\Ix}           {\ensuremath{\overline{\Xi}^+}\xspace}
\newcommand{\Xis}          {\ensuremath{\Xi^{\pm}}\xspace}
\newcommand{\Oms}          {\ensuremath{\Omega^{\pm}}\xspace}
\newcommand{\degree}       {\ensuremath{^{\rm o}}\xspace}
\newcommand{\jpsi} {\ensuremath{{\mathrm J}/\psi}\xspace}
\newcommand{\psip} {\ensuremath{\psi'}\xspace}
\newcommand{\rhozero} {\ensuremath{\rho^0}\xspace}
\newcommand{\Chisq}{\ensuremath{\chi^{2}}}
\newcommand{\twohundrednn}       {$\sqrt{s_{\mathrm{NN}}}~=~200$~Ge\kern-.1emV\xspace}
\newcommand{\AD}   {\rm{AD}\xspace}
\newcommand{\ADA}  {\rm{ADA}\xspace}
\newcommand{\ADC}  {\rm{ADC}\xspace}
\newcommand{\fI}   {f_{\rm I}}
\newcommand{\fd}   {f_{\rm D}}
\newcommand\todo[1]{\textcolor{red}{#1}}
\newcommand{\starlight}{STARlight\xspace}
\newcommand{\montecarlo}{Monte Carlo\xspace}
\newcommand{\AxE}{\ensuremath{A\times{\epsilon}}}
\newcommand{\lambdatheta}{\ensuremath{\lambda_\theta}}
\newcommand{\lambdaphi}{\ensuremath{\lambda_\varphi}}
\newcommand{\lambdathetaphi}{\ensuremath{\lambda_{\theta\varphi}}}
\newcommand{\tildephi}{\ensuremath{\widetilde{\varphi}}}
%%%%%%%%%%%%%%%  Title page %%%%%%%%%%%%%%%%%%%%%%%%
\begin{titlepage}
% the dates below correspond to CERN approvals
% please don't touch: EB chairs will take care
\PHyear{2023}       % required, will be obtained from CERN
\PHnumber{068}      % required, will be obtained from CERN
\PHdate{19 April}  % required, will be obtained from CERN
%%%%%%%%%%%%%%%%%%%%%%%%%%%%%%%%%%%%%%%%%%%%%%%%%%%%

%%% Put your own title + short title here:
\title{First polarisation measurement of coherently photoproduced \jpsi in ultra-peripheral \PbPb collisions at \fivenn
}
\ShortTitle{\jpsi polarisation in Pb--Pb UPC with the ALICE experiment}   % appears on left page headers

%%% Do not change the next lines
\Collaboration{ALICE Collaboration\thanks{See Appendix~\ref{app:collab} for the list of collaboration members}}
\ShortAuthor{ALICE Collaboration} % appears on right page headers, do not change

\begin{abstract}
The first measurement of the polarisation of coherently photoproduced \jpsi mesons in ultra-peripheral \PbPb collisions, using data at \fivenn, is presented. The \jpsi meson is measured via its dimuon decay channel in the forward rapidity interval $-4.0 < y < -2.5$ using the ALICE detector at the CERN LHC. An event sample corresponding to an integrated luminosity of 750~$\mu\text{b}^{-1}$ $\pm$ 5\% (syst) is analysed. Hadronic activity is highly suppressed since the interaction is mediated by a photon. The polar and azimuthal angle distributions of the decay muons are measured, and the polarisation parameters $\mathbf{\lambda_{\theta}}$,  $\mathbf{\lambda_{\varphi}}$, $\mathbf{\lambda_{\theta\varphi}}$ are extracted. The analysis is carried out in the helicity frame. The results are found to be consistent with a transversely polarised \jpsi. These values are compared with previous measurements by the H1 and ZEUS experiments. The polarisation parameters of coherent \jpsi photoproduction in \PbPb collisions are found to be consistent with the $s$-channel helicity conservation hypothesis.

\end{abstract}
\end{titlepage}

\setcounter{page}{2} %please do not remove this line

%%%%%%%%%%%%%%%%%%%%%%%%%%%%%%%%
% begin main text
%%%%%%%%%%%%%%%%%%%%%%%%%%%%%%%%

\section{Introduction}
The study of vector meson photoproduction at high energies has received great interest in recent years~\cite{Baltz:2007kq,Contreras:2015dqa,Klein:2020fmr}.
Ultra-peripheral collisions (UPCs) at the CERN Large Hadron Collider (LHC) allow access to the highest available photon energies for such photon--induced interactions.
In UPCs, the impact parameter is larger than the sum of the radii of the two colliding nuclei, so they are beyond the range of the nuclear force and do not interact hadronically.
A quasi-real photon is emitted from one of the charged nuclei and this in turn might fluctuate into a quark--antiquark pair, which can scatter elastically off the other nucleus. Vector meson photoproduction off nuclei can be either coherent, with the photon coupling coherently to the nucleus as a whole, or incoherent, where instead the photon couples to a single nucleon.
Much emphasis has been put on how this process could give access to space and momentum features of the momentum distribution of gluons of the hadronic target. In particular, several model calculations indicate that vector meson photoproduction is sensitive to nuclear modifications of the partonic structure for gluons that carry a rather small fraction of the momentum of the  nucleons~\cite{Guzey:2016piu,Bendova:2020hbb}. 
Kinematics dictates that the momentum fraction carried by the gluons is given by $x=\frac{M_{\jpsi}}{\snn}\, e^{\pm y}$, where $M_{\jpsi}$ is the mass of the \jpsi, $y$ is the rapidity, and \snn is the centre-of-mass rapidity per nucleon--nucleon collision. For rapidities $y \neq 0$ one thus has a two-fold ambiguity in the value of $x$. In the middle of the acceptance of the muon spectrometer $y = -3.25$. This corresponds to $x$-values of $2\times 10^{-3}$ and $5\times 10^{-5}$. The largest contribution to the cross section comes from the higher value.
Recent experimental data from ALICE, CMS, and LHCb~\cite{Abelev:2012ba,Abbas:2013oua,Khachatryan:2016qhq,Tumasyan_2023,LHCb:2021bfl} have found experimental evidence of gluon shadowing~\cite{Guzey:2013xba}.

In previous analyses of exclusive vector meson photoproduction off protons at lower energies, using both real and virtual photons, the vector meson has generally been found to have transverse polarisation, as would be expected if the vector meson keeps the polarisation of the incoming photon ($s$-channel helicity conservation, SCHC)~\cite{Gilman:1970vi, Schilling:1973ag}. However, although early results from SLAC~\cite{Ballam:1972eq} and from the CERN SPS~\cite{OMEGAPhoton:1991exy}  with $\rho^0$ mesons produced from real photons are consistent with SCHC, a much more recent study of \rhozero, $\omega$, and $\phi$ photoproduction from CLAS~\cite{CLAS:2008rpm,Vernarsky:2014vaa,Dey:2014tfa} reports violation of SCHC.

Exclusive photoproduction of vector mesons has also been studied at HERA in ep collisions. Both the H1~\cite{Aktas:2005xu} and ZEUS~\cite{Chekanov:2004mw} collaborations observed transverse polarisation in \jpsi photoproduction, thus favouring SCHC, in their photoproduction data. Note that at HERA, photoproduction, at very low photon $Q^2$ ($Q^2 \lesssim 0.05 $ \GeVtwo), is  distinguished from electroproduction at somewhat higher $Q^2$, even though both are photon--hadron interactions. Here $Q^{2}$ is the four-momentum transfer squared from the lepton to the nucleon vertex. In electroproduction the photon is more  virtual, and can be partially longitudinally polarised. Accordingly, in \jpsi electroproduction the degree of longitudinal polarisation is seen to increase with $Q^2$~\cite{ZEUS:1998fem}. Photoproduction can also be studied in heavy-ion UPCs. Polarisation measurements have been reported by STAR for $\rho^0$ photoproduction~\cite{Abelev_2008} in collisions of gold nuclei at RHIC, and are consistent with SCHC.

For all previous vector meson photoproduction analyses at the LHC,  when calculating the experimental acceptance, the vector meson has been assumed to have the same helicity as that of the initial photon that interacted with the target, i.e. SCHC.
Since there is no fundamental physics principle that justifies SCHC, new polarisation and spin parameter measurements of photoproduced vector mesons at high energies could provide important insights.

In this Letter, the first polarisation measurement of coherent \jpsi photoproduction in ultra-peripheral Pb--Pb collisions at \fivenn is presented. These events are defined by the reaction $\gamma + \rm{Pb}\rightarrow \jpsi + Pb$, where the photon is emitted from one of the two colliding lead nuclei, and the \jpsi meson is reconstructed in the forward rapidity interval $-4.0 < y < -2.5$.

\section{Detector description}
The ALICE detector and its performance are described in~\cite{Aamodt:2008zz,Abelev:2014ffa}. Muons from \jpsi decays are measured in the single-arm muon spectrometer. For coherent \jpsi photoproduction, the hadron activity is required to be highly suppressed. For this reason, vetoes are applied using the \VZERO and ALICE Diffractive (\AD) detectors. The muon spectrometer covers the pseudorapidity interval $-4.0<\eta<-2.5$. It consists of a ten interaction length absorber followed by five tracking stations, the third of which is placed inside a dipole magnet with a 3 Tm integrated magnetic field, a 7.2 interaction length iron wall, and a trigger system located downstream of the iron wall. Each tracking station is made of two planes of cathode pad chambers, while the trigger system consists of four planes of resistive plate chambers arranged in two stations. Muon tracks are reconstructed using the tracking algorithm described in~\cite{Aamodt:2011gj}.
The \VZERO detector is composed of the \VZEROA and \VZEROC sub-detectors, consisting of 32 cells each and covering the pseudorapidity intervals $2.8< \eta <5.1$ and $-3.7<\eta <-1.7$, respectively. The AD detector~\cite{Broz:2020ejr} is composed of the \ADC and \ADA sub-detectors with 4 cells each, located at $-19.5$ m and $+16.9$ m along the beam direction from the interaction point and covering the pseudorapidity intervals $-7.0 < \eta < -4.9$ and $4.7 <\eta < 6.3$, respectively~\cite{N.Cartiglia:2015gve}. The \VZERO and \AD detectors are scintillator arrays with a time resolution better than 1 ns, allowing a distinction between beam--beam and beam--gas interactions.

\section{Data selection and analysis}
This analysis follows the same  selections as those recently described in~\cite{Acharya:2019vlb}, which combines the ultra-peripheral \PbPb datasets at  centre-of-mass energy per nucleon--nucleon collision \fivenn collected in 2015 and 2018. The resulting dataset corresponds to an integrated luminosity of about 750~$\mu\text{b}^{-1}$ $\pm$ 5\% (syst). The UPC event sample was recorded using a hardware trigger that requires at least two unlike-sign tracks in the muon spectrometer, and no activity in the \VZEROA,  \ADA, and \ADC detectors. The use of these forward detectors ensures the exclusivity condition. The single muon threshold for the transverse momentum {\pt} for the muon trigger system has been set to \pt $ = 1\text{ GeV/}c$~\cite{Bossu:2012jt}. At the offline level further selections are applied. First, exactly two unlike-sign tracks are required in the muon spectrometer, where both tracks must be compatible with high-quality reconstructed muons as described in~\cite{Acharya:2019vlb}. The tracks must fulfil the requirement on the radial coordinate of the track at the end of the absorber, and on the extrapolation to the nominal vertex~\cite{ALICE:2012yye}. Track segments in the tracking chambers are matched with corresponding segments in the trigger chambers. In addition, each muon and each dimuon is required to be in the pseudorapidity $-4.0 < \eta < -2.5$ and rapidity $-4.0 < y < -2.5$ intervals, respectively. In order to select  events where coherent \jpsi production is strongly enhanced, the dimuon {\pt} is required to be less than $0.25\text{ GeV/}c$, effectively rejecting most of the non--coherent \jpsi events as discussed in~\cite{Acharya:2019vlb}. Finally, the \VZEROA, \ADA, and \ADC vetoes are applied, and up to two hits are allowed in \VZEROC, since the \VZEROC acceptance overlaps with the that of the muon spectrometer. These offline level selections have wider time windows than those used for the trigger electronics, together with a more refined algorithm, allowing for a more precise rejection. Possible \VZERO and \AD detector veto inefficiencies could originate from independent hadronic or electromagnetic pile-up interactions. The mean efficiency correction factor is 95\%, accounting for \VZEROA and \AD veto inefficiencies, while the \VZEROC veto is found to be fully efficient. Since the angular distributions studied in this analysis are not affected by the forward veto selections, only the muon-related efficiencies are considered.

The detector acceptance and efficiency (\AxE) for \jpsi and for \psip are evaluated using a sample of events generated with \starlight 2.2.0~\cite{Klein:2016yzr}, which models the \jpsi as transversely polarised. The decay muons are then propagated using a model of the detector implemented in GEANT 3.21~\cite{Brun:1082634} with realistic conditions. They are then reconstructed with the same algorithm as that used for the experimental data. This sample is also used to construct the response matrices which are needed for this measurement. Typical values of the \AxE\ range from 0.41 for $\cos\theta \sim 0$, and down to 0.024 at the limit of the acceptance i.e.  $\cos\theta \sim 0.56$ (see next section for definition of $\theta$).

\section{Angular distributions}
The angular distribution (W) of the decay muons is described in terms of the three polarisation parameters \lambdatheta, \lambdaphi, and \lambdathetaphi\ using the following expression as a function of the polar angle $\theta$ and of the azimuthal angle $\varphi$, which respects parity conservation~\cite{Faccioli:2010kd}:
\begin{equation}
  W(\cos\theta, \varphi) \propto \frac{1}{3+\lambda_\theta}\,\left[ 1+ \lambda_\theta \, \cos^2\theta + \lambda_\varphi\,\sin^2\theta\,\cos2\varphi + \lambda_{\theta\varphi}\,\sin2\theta \, \cos\varphi \right]\text{.}
  \label{eq2}
\end{equation}
It is also possible to study the data by using the following expressions
\begin{equation}
\begin{split}
W(\cos\theta)                  & \propto \frac{1}{3 + \lambda_\theta}\, \left[ 1 + \lambda_\theta \, \cos^2\theta \right]\text{ ,} \\
W(\varphi)                        & \propto  1 + \frac{2\lambda_\varphi}{3 + \lambda_\theta} \, \cos2\varphi \text{ ,} \\
\end{split}
\label{eq:split}
\end{equation}
obtained by integrating Eq.~\ref{eq2} in the corresponding angular variable.

The polarisation parameters quantify the degree of polarisation. When all parameters are consistent with zero, an isotropic angular distribution is obtained. When $(\lambda_\theta, \lambda_\varphi, \lambda_{\theta\varphi}) = (1, 0, 0)$ the vector meson is transversely polarised, while $(\lambda_\theta, \lambda_\varphi, \lambda_{\theta\varphi}) = (-1, 0, 0)$ corresponds to a purely longitudinal polarisation.

The angular distribution of the decay muons is studied using the helicity frame. This frame has been used in recent studies of the polarisation of inclusive \jpsi production in \pp and \PbPb collisions by the ALICE Collaboration~\cite{Abelev:2011md,Acharya:2018uww, Acharya:2020xko, ALICE:2022sli}, and by the PHENIX Collaboration in \pp collisions~\cite{Acharya:2020oym}. In the helicity frame the $z$-axis coincides with the flight direction of the \jpsi in the \PbPb centre-of-mass frame. The $y$-axis is perpendicular to the plane formed by the collision axis and the direction of the \jpsi in the \PbPb centre-of-mass frame, while the $x$-axis is chosen so as to have a right-handed triplet~\cite{Faccioli:2010ji}. The angles refer to the direction of the positive decay muon in this frame. Another commonly used definition of the axes is the Collins--Soper frame~\cite{Faccioli:2010kd}. For \jpsi events at low \pt both frames are compatible, so results are only reported in  terms of the helicity frame. The use of different reference frames has also been justified in terms of how the background could behave differently than the signal in the various frames. As discussed below, the background in this analysis is small which is expected for coherent \jpsi events.

\begin{figure}[ht!]\
     \begin{center}
        \subfigure[]{
            \label{fig:SigEx}
            \includegraphics[width=0.7\textwidth]{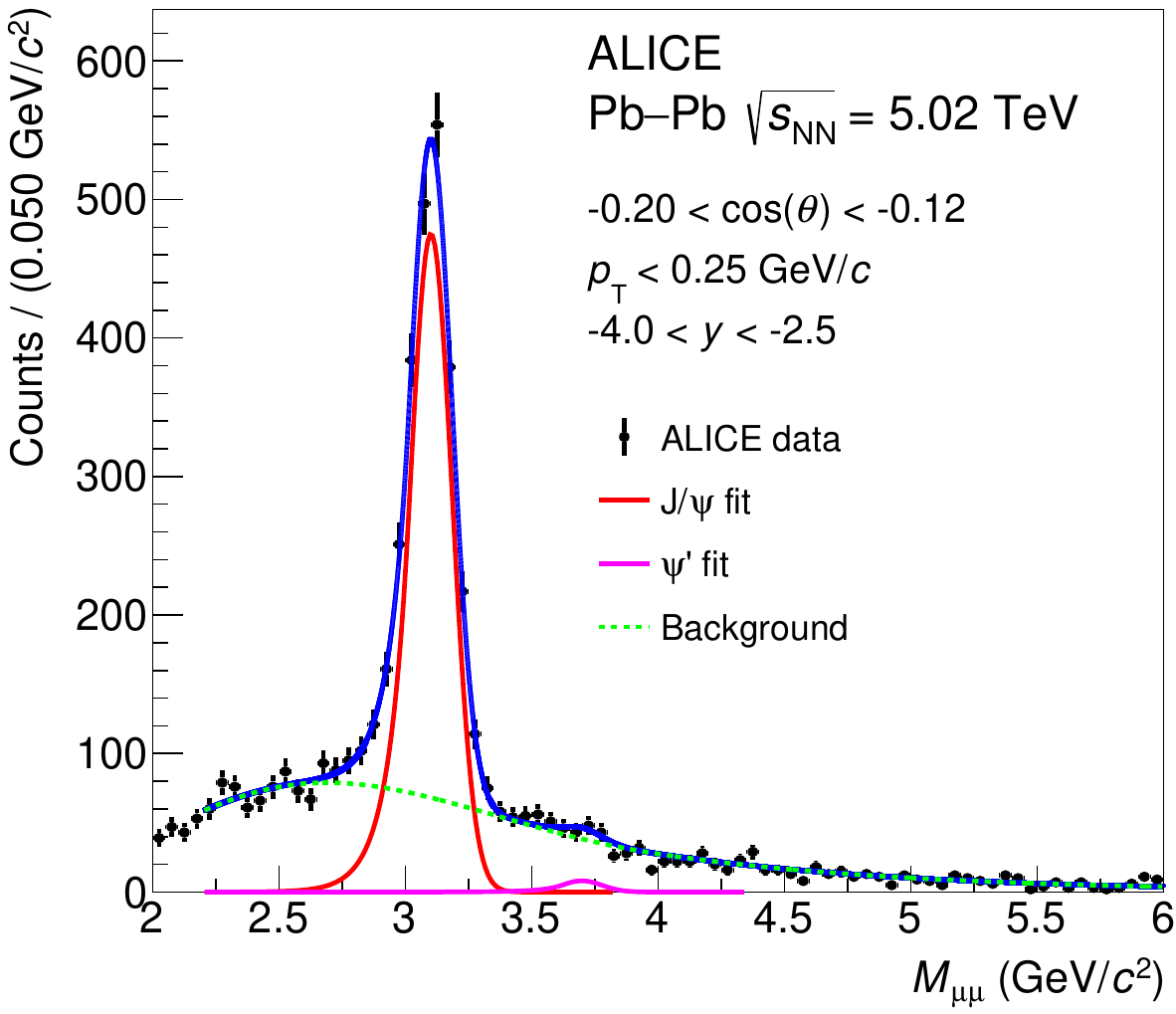}
        }
        \subfigure[]{
            \label{fig:SigPt}
            \includegraphics[width=0.7\textwidth]{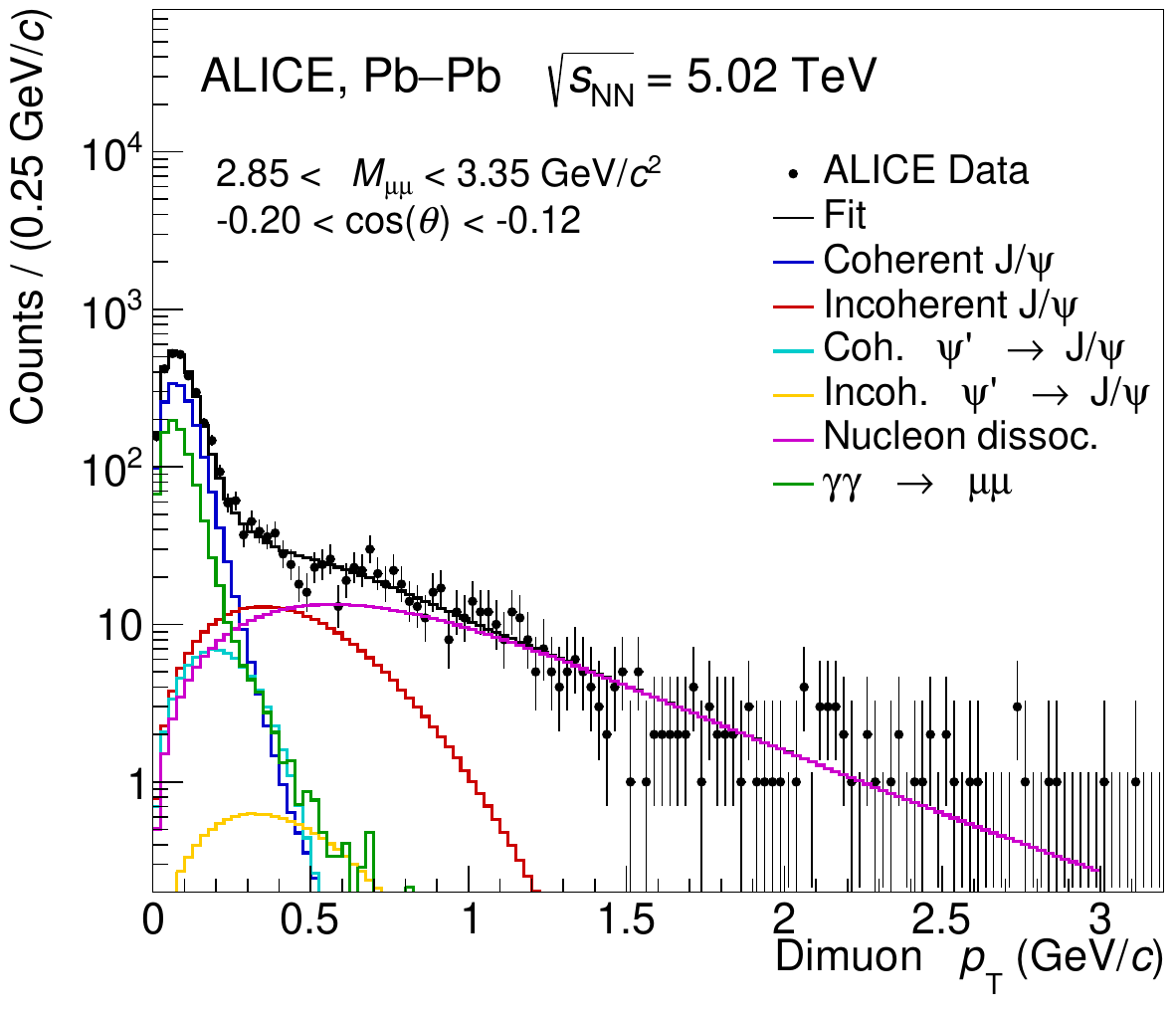}
        }
      \end{center}
  \caption{A fit to the raw invariant mass (a) and \pt distributions for events in the $-0.20 < \cos\theta < -0.12$ range evaluated in the helicity frame. The bottom panel (b) shows the dimuon \pt around the \jpsi mass range.
  The displayed uncertainties are statistical.}
  \label{fig:Signal}
\end{figure}

To extract the \jpsi yield, the dimuon invariant mass distribution is fitted in the interval $2.2 < M_{\mu\mu} < 6.0\text{ GeV/}c^2$, requiring in addition that $\pt<0.25$~\GeVc to enhance the contribution from coherent production. The study is performed in twenty-four bins of $\cos\theta$ over the interval $-1<\cos\theta < 1$, and six bins in $\varphi$ in the interval between 0 and $2 \pi$, to obtain yields in intervals of $\cos\theta$ and $\varphi$. As an example, Fig.~\ref{fig:SigEx} shows the invariant mass distribution of coherent \jpsi candidates in the interval  $-0.20 < \cos{\theta} < -0.12$. We follow the  procedure described in~\cite{Acharya:2019vlb} for fitting the invariant mass distributions. The data are fitted to a background model (dimuon continuum) and to a signal model (\jpsi and \psip resonances). For the background, the fitting function is given by a fourth-order polynomial multiplied by an exponential. The exponential component becomes important for $M_{\mu\mu} > 4.0 \text{ GeV/}c^2$. The coefficients of the polynomial function are fixed to those obtained from Monte Carlo simulations using the \starlight~\cite{Klein:2016yzr} generator for the $\gamma\gamma \longrightarrow \mu^+\mu^-$ process. The exponential tail behaviour is expected for such an electromagnetic process. At low dimuon masses the detector acceptance decreases due to the trigger requirement, so the exponential part is not needed and a simple fourth-order polynomial is used. The \jpsi and \psip signals are fitted with one-sided Crystal Ball functions~\cite{Gaiser:1982yw}. The parameters for the Crystal Ball functions are fixed to those obtained from Monte Carlo templates. The mean and width of the \jpsi peak are free parameters in the fit. The mean of the \psip peak is fixed to the world average value~\cite{Tanabashi:2018oca}\   and its width to that of the {\jpsi} multiplied by the ratio of the widths of the \psip to the \jpsi, as measured from the analysis of reconstructed \starlight events. The ratio is approximately 1.09.

The \jpsi yield extracted from the fit to the invariant mass distributions comprises five contributions, namely ({\it i}) coherent production, ({\it ii}) incoherent production, and ({\it iii}) feed-down \jpsi production from coherent \psip production,  incoherent \psip production, and ({\it v}) incoherent \jpsi production with associated nucleon production. To extract the coherent \jpsi yield, the residual background has been estimated from fits to the dimuon \pt distribution. This is possible since the various processes have different momentum distributions of the produced \jpsi mesons. Incoherent \jpsi photoproduction accompanied by nucleon dissociation was also taken into account in order to describe the high \pt tail, using a template based on the H1 parametrization of the
 dissociative \jpsi photoproduction \cite{Alexa:2013xxa}. Following this approach, which has been described in a recent publication~\cite{Acharya:2019vlb}, the background contribution is found to be smaller than 4.6\% for the integrated $\cos{\theta}$ and $\varphi$ intervals used in this analysis. Note this fraction does not include the $\gamma\gamma \longrightarrow \mu^+\mu^-$ background which has been subtracted from fits to the invariant mass distribution, as described above. Since the residual background is small, and assuming that those \jpsi events have the same polarisation to that of coherent \jpsi events, their contributions are not subtracted from the sample. The \jpsi yield is estimated from the dimuon invariant mass distribution in a number of ($\cos(\theta),\varphi)$ cells assuming it all to be coherent. 
For illustration purposes, Figure~\ref{fig:SigPt} shows the \pt distribution of candidates in the \jpsi invariant mass peak region ($2.85<M_{\mu\mu}<3.35\text{ GeV/}c^2$) and in the same $\cos(\theta)$ interval of Fig.~\ref{fig:SigEx}.

\begin{figure}[ht!]
    \begin{center}
    \includegraphics[width = 0.7\textwidth]{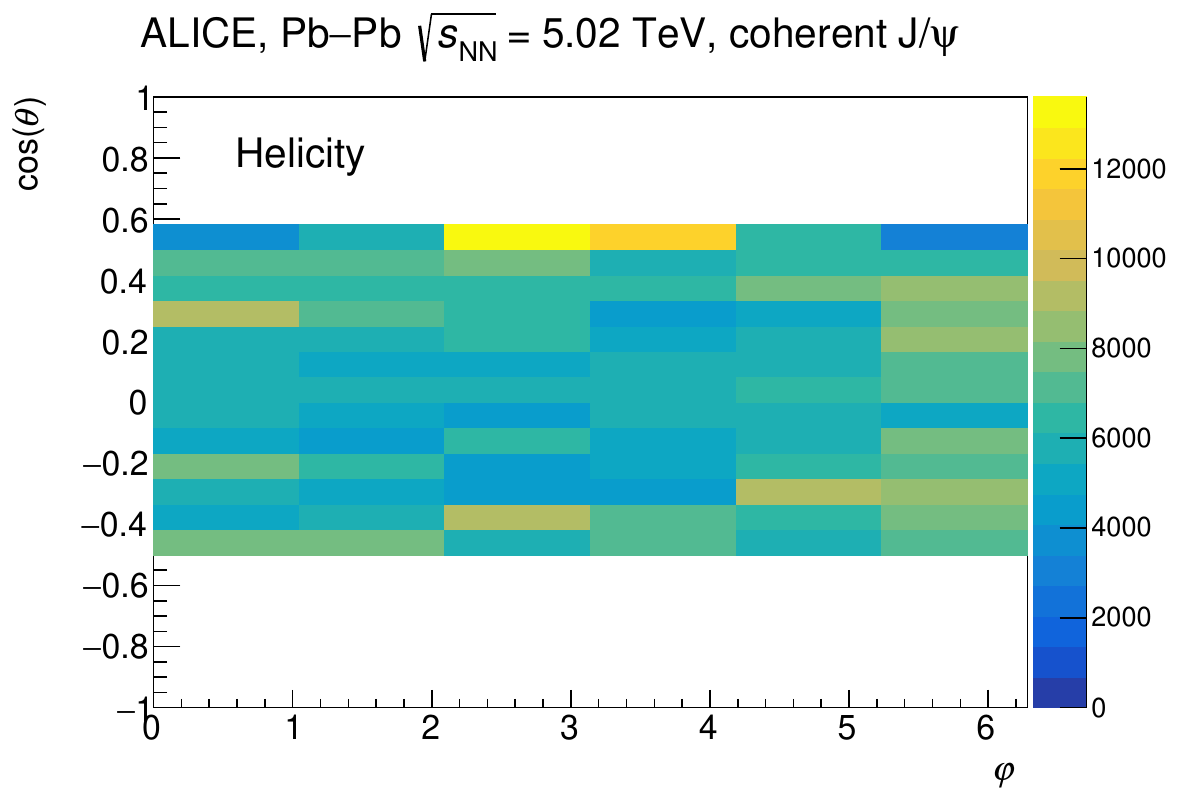}
    \end{center}
  \caption{Two-dimensional angular distribution of coherent $\jpsi$ candidates in the helicity frame, after being unfolded in $\varphi$ and corrected for \AxE\ in $\cos\theta$. }
  \label{fig:2DHE}
\end{figure}

To correct for detector effects, the data are unfolded and corrected for \AxE\ in cells of $\cos \theta$ and $\varphi$, and in the 1D distributions,  \AxE\ corrections are applied to the data in $\cos\theta$. Since the STARLight Monte Carlo has been found to give a reasonable description of the data, in particular for coherent \jpsi production (low \pt events) at forward rapidity, we take this to be a natural starting point. The unfolding uses the Bayes' iterative method~\cite{DAgostini:1994fjx} as implemented in the RooUnfold framework~\cite{Adye:2011gm}. The number of iterations in the unfolding procedure was set to fifteen. This number was found to be optimal at minimising the average uncertainty, following the prescription discussed in~\cite{Schmitt:2016orm}. 
Since the generated $\cos \theta$ is recovered in  the reconstructed $\cos \theta$ with a spread smaller than the bin width, bin migration effects are negligible and the data do not need unfolding in $\cos \theta$. Therefore, as in addition the sample size is small, the unfolding is performed only for $\varphi$, as described below. Figure~\ref{fig:2DHE} shows the corrected \jpsi distribution in $\cos \theta$ and $\varphi$, where the $\varphi$ distribution in addition has been unfolded. The unfolding procedure increases the uncertainties on the estimated number of \jpsi{s} in each cell significantly. The first step is to fit the $\cos \theta$ distribution as a 1D distribution, treating the uncertainties in each bin as independent of those of the other bins.
The $\cos\theta$ distribution is fitted using a $\chi^{2}$ fitting method and  the equivalent of Eq.\ (\ref{eq:split}) with the normalisation modified to take into account the actual limits of integration, $-0.5<\cos\theta<0.583$. This gives $\lambdatheta = 0.93 \pm 0.05\mathrm{(stat)} \pm 0.22 \mathrm{(sys)}$.

For the $\varphi$-related parameters \lambdaphi\ and \lambdathetaphi, the 2D distribution is required, giving the best estimates of the content of each $(\cos\theta,\varphi)$ cell, where the $\varphi$ distribution is unfolded independently in each $\cos\theta$ band. As the contents of the unfolded cells for a given $\cos\theta$ are correlated, an (input) covariance matrix was determined separately for each $\cos\theta$ band. The contents of each cell and their uncertainty, as determined before any unfolding from the fit to the invariant mass distribution, are used to generate 10000 samples with fluctuating contents according to a normal distribution. The contents of all bins are varied simultaneously, and the unfolding algorithm applied. The covariance matrix for the cell contents after unfolding is evaluated using $\mathrm{cov}(x,y) = <xy> -<x><y>$. Finally, the whole (2D) angular distribution is fitted with Eq. (\ref{eq2}), using a $\chi^{2}$ method, the unfolded contents for the central values in each cell, and the set of convariance matrices. 
A free fit for all three parameters yields $\lambdatheta = 0.82 \pm 0.44$,\ $\lambdaphi = 0.03 \pm 0.04$, and $\lambdathetaphi = 0.10 \pm 0.09$. However, for the final fit the $\lambdatheta$ value has been fixed to the value obtained from the 1D fit, allowing only $\lambdaphi$ and $\lambdathetaphi$ to vary. The $\Chisq/NDF$ for this final fit is $114.2/175$.

Table~\ref{tab:HE} presents a summary of the various sources of systematic uncertainty on the measurement of the polarisation parameters for coherent \jpsi mesons. To study how the polarisation parameters vary as the acceptance changes when $\cos\theta$ increases, the polar angle distribution is fitted in the  $\cos\theta$ ranges $[-0.5, 0.58{3}]$ and $[-0.5, 0.5]$, as the analysis is carried out in a restricted $\cos\theta$ interval. The observed variation is taken as a systematic uncertainty. The uncertainty from the signal extraction is estimated by changing the \jpsi line shape from a single-sided Crystal Ball function to a double-sided one. The background fraction is thus adjusted accordingly in the fit. The uncertainty from the unfolding is assessed by varying the number of iterations by $\pm 1$ with respect to the number chosen for the central value. The impact of the response matrix as a source of systematic uncertainty in the results is also investigated. Three different matrices corresponding to transverse polarisation, flat distribution and longitudinal polarisation have been used, and the differences in the fitted parameters were evaluated to determine the corresponding systematic uncertainty. Another source of systematic uncertainty is the trigger and tracking efficiency. This is studied by varying the offline \pt selection from 0.85 to 1.15 GeV/$c$ for single muons, corresponding to a $\pm 5\%$ variation on the trigger efficiency.  The choice of \pt selection and the $\cos\theta$ range are correlated. The larger the \pt, the smaller the allowed $\cos\theta$ range. Finally, the global systematic uncertainty on the polarisation parameters is obtained by adding the results of the various sources in quadrature.

\begin{table}[!htb]
\centering
\caption{Summary of the systematic uncertainty contributions, presented as absolute values. The $\cos\theta$ range systematic uncertainty refers to the fitted range variation, the signal extraction to the choice of the description of the \jpsi, the unfolding systematic uncertainty is due to the choice of the number of iterations, the response matrix refers to the input distribution in generating the matrix, and the trigger systematic uncertainty is associated to the single muon \pt selection used for the trigger efficiency calculation.}
\hspace{1.5 mm}
\label{tab:HE}
\begin{tabular}{|l|l||lll|}
    \hline
    & 1D &\multicolumn{3}{c|}{2D}\\
	Systematics & \lambdatheta & \lambdatheta & \lambdaphi  & \lambdathetaphi   \\
	\hline
	\hline
	$\cos\theta$ range         &	0.079			& 0.142 & 0.002 & 0.056 \\
	signal extraction          &	0.002			& 0.026 & 0.002 & 0.008 \\
	unfolding	               &    0.0     		& 0.019 & 0.004 & 0.004  \\
	response matrix            &	0.010			& 0.009 & 0.008 & 0.004 \\
	single muon \pt threshold  &    0.203           & 0.196 & 0.022 & 0.019 \\
	\hline
	Total  					   & 	0.218           & 0.244 & 0.023 & 0.060 \\	\hline	
\end{tabular}
\end{table}

\section{Results and discussion}

\begin{figure}[t!]
     \begin{center}
        \subfigure[]{
            \label{fig:CosThetaCsMinuit}
            \includegraphics[width=0.6\textwidth]{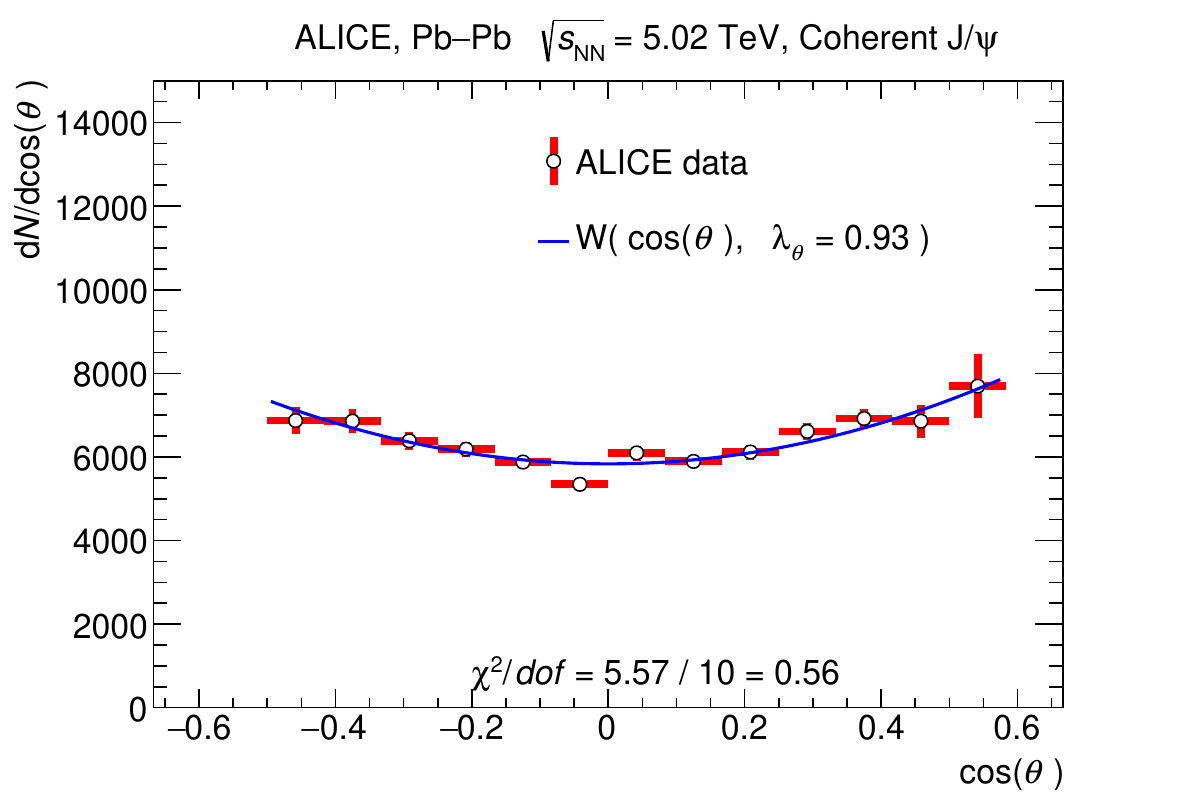}
        }\\
        \subfigure[]{
            \label{fig:PhiCsMinuit}
            \includegraphics[width=0.6\textwidth]{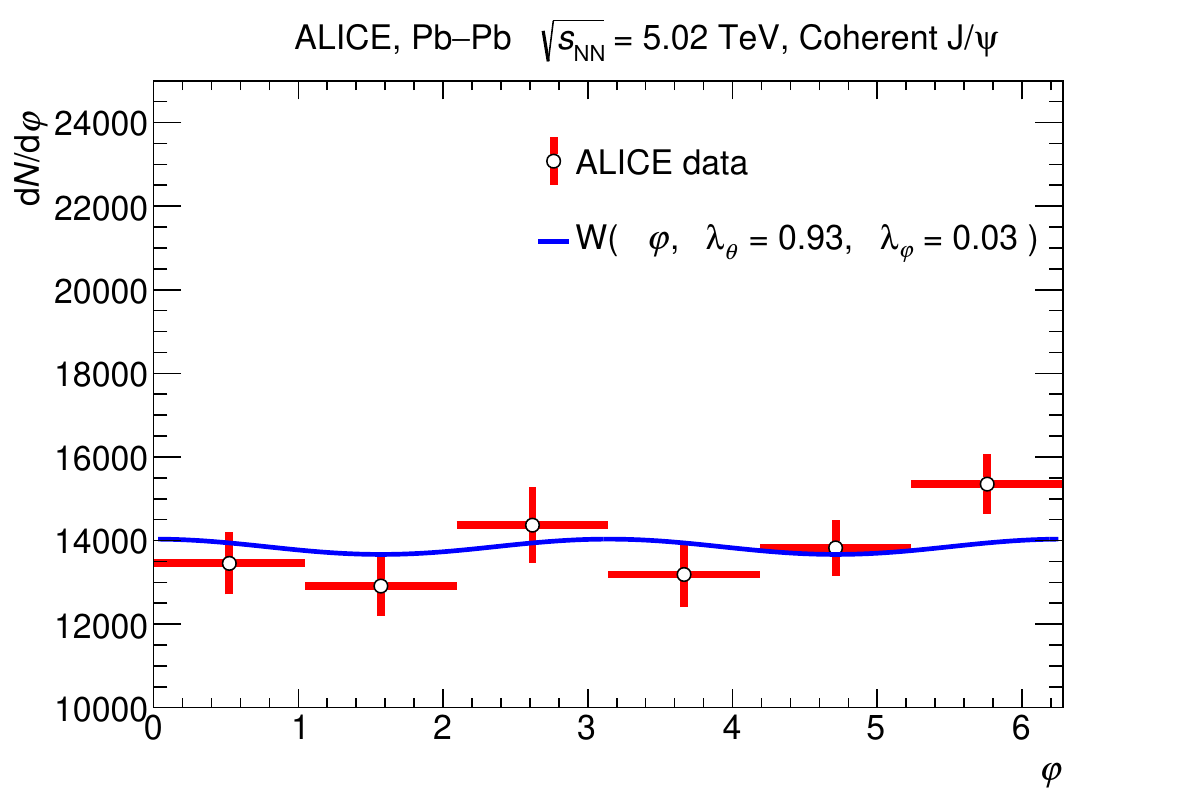}
        }\\
      \end{center}
  \caption{Projections of the coherent \jpsi data points in (a) $\cos\theta$ and (b) $\varphi$, for events in the $-4 < y < -2.5$ rapidity interval. The curves are obtained by setting the lambda parameters in Eq.~\ref{eq:split} to the results of the fit to the distribution given  in Table~\ref{tab:summary}. The uncertainties on the data points are statistical.}
  \label{fig:1DresultsCS}
\end{figure}

The projections of the angular distribution of the yield of coherently produced \jpsi in terms of $\cos\theta$ and $\varphi$ are shown in
Figs.~\ref{fig:CosThetaCsMinuit} and~\ref{fig:PhiCsMinuit}, respectively. The numerical values of the polarisation parameters are given in Table~\ref{tab:summary}. The numerical values of the polarisation parameters are given in Table 2, with the statistical uncertainty defined using $\Delta\chi^{2} = 2.3$, since this is a two-dimensional distribution\cite{Drijard:318006,ParticleDataGroup:2024cfk}, and the systematic error calculated as described above.  Note that the $\lambdatheta$ result is taken from the 1D fit. The results are found to be consistent with $(\lambda_\theta, \lambda_\varphi, \lambda_{\theta\varphi}) = (1, 0, 0) $, i.e. transverse polarisation, and they satisfy the required positivity constraints~\cite{Martens:2017cvj}.

\begin{table}[bt!]
\centering
\caption{Measured polarisation parameters for coherent \jpsi in the helicity  frame. The central values are given with statistical and total systematic uncertainties.}

\label{tab:summary}
	\hspace{1.5 mm}

	\begin{tabular}{|c|c|c|}

		\hline
		\textbf{\lambdatheta} & \textbf{\lambdaphi} & \textbf{\lambdathetaphi} \\ \hline
		 $0.93 \pm 0.05 \pm 0.22$ & $0.03 \pm 0.04 \pm 0.02$ & $0.10 \pm 0.09 \pm 0.06 $\\ \hline
	\end{tabular}
\end{table}

\begin{figure}[ht!]
    \begin{center}
    \includegraphics[width = 0.7\textwidth]{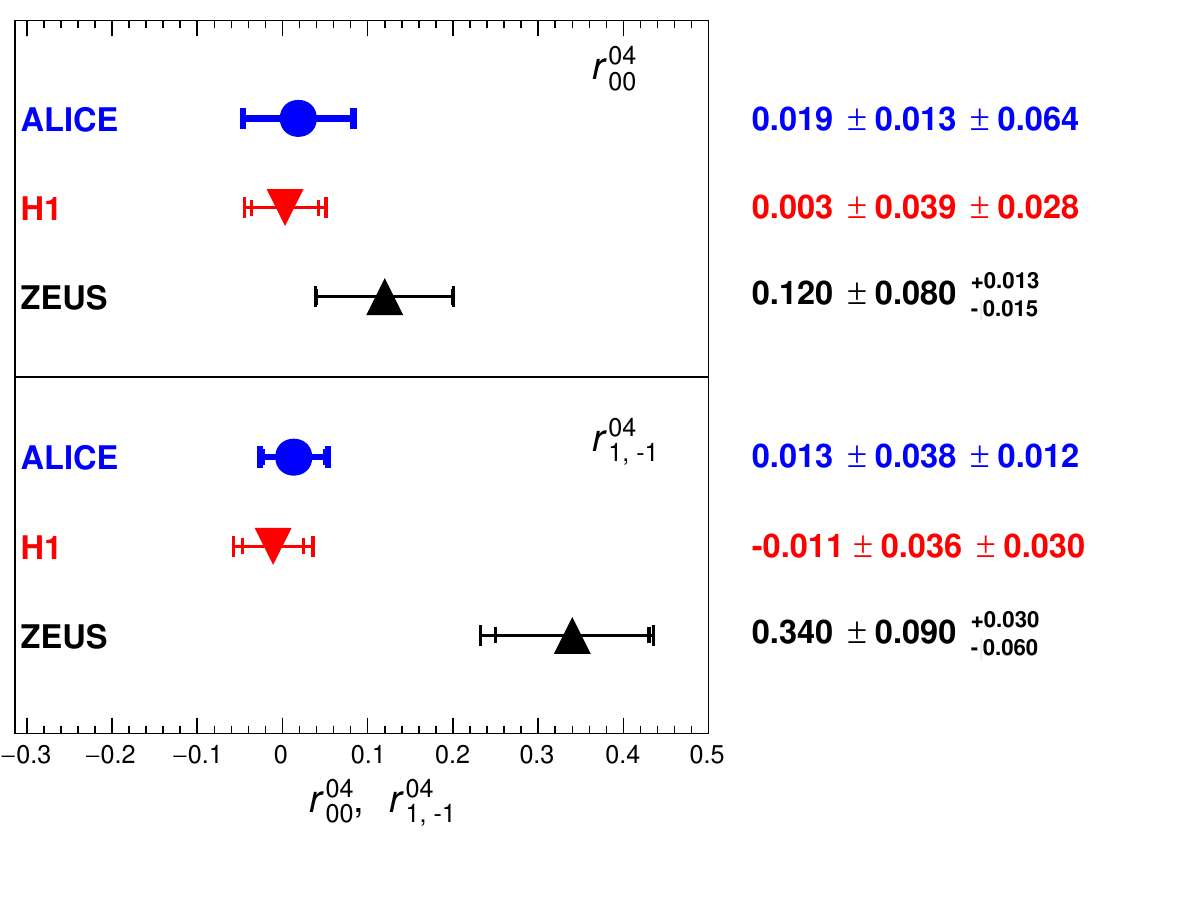}
    \end{center}
  \caption{ALICE results displayed in terms of spin density matrix elements, along with statistical and systematic uncertainties. A comparison with the available lowest $Q^2$ ranges of the results from the  H1 and ZEUS collaborations is also provided. H1 results~\cite{Aktas:2005xu} correspond to $Q^2 \sim 0.05$ \GeVtwo consistent with photoproduction, while the shown ZEUS measurement~\cite{Chekanov:2004mw} is for $2 < Q^2 < 5$ \GeVtwo. }
  \label{fig:H1ZEUS}
\end{figure}

Alternative formalisms exist for describing the polarisation parameters~\cite{Schilling:1973ag,Martens:2017cvj,Gavrilova:2019jea}. In order to provide a comparison with H1~\cite{Aktas:2005xu} and ZEUS~\cite{Chekanov:2004mw} results, the measured $\mathbf{\lambda_{\theta}}$ and $\mathbf{\lambda_{\varphi}}$ values are converted to spin-density matrix elements $r_{\lambda\lambda^\prime}^{\rm ik}$~\cite{Schilling:1973ag}.
The conversion from lambda parameters to spin-density matrix elements $r_{\lambda\lambda^\prime}^{\rm ik}$ can be obtained from the following expressions~\cite{Aktas:2005xu}:
\begin{equation}
\begin{split}
W(\cos\theta) & \propto 1 + r_{00}^{04} + (1 - 3r_{00}^{04})\, \cos^2\theta \\
W(\varphi)    & \propto 1 +  r_{1,-1}^{04} \, \cos2\varphi\text{ .}
\end{split}
\label{eq:splitHERA}
\end{equation}
By comparing Eq.~\ref{eq:split} and Eq.~\ref{eq:splitHERA}, the spin-density matrix elements are expressed as:
\begin{equation}
\begin{split}
r_{00}^{04}   & = \frac{1 - \lambda_\theta}{3 + \lambda_\theta} \\
r_{1,-1}^{04} & = \frac{\lambda_\varphi}{2}\,(1 + r_{00}^{04})\text{ .}
\end{split}
\label{eq:spin}
\end{equation}
Figure~\ref{fig:H1ZEUS} shows the results of the comparison. In this analysis, the HERA measurements are given in  the helicity frame. The ALICE result presented in Figure~\ref{fig:H1ZEUS} is compatible with the H1 measurement, which also uses a photoproduction sample ($Q^2 \lesssim 0.05$ \GeVtwo), and is lower than the ZEUS measurement, which uses a sample with $2 < Q^2 < 5$~\GeVtwo~\cite{Chekanov:2004mw}. This is the lowest $Q^{2}$ point in the ZEUS range, and 
it can be seen to be tending towards the photoproduction value found by H1, but is already  in the electroproduction regime.

\section{Summary}
The first measurement of the polarisation of coherent \jpsi photoproduction in ultra-peripheral \PbPb collisions, using data at \fivenn, has been carried out by the ALICE Collaboration. The $\mathbf{\lambda_{\theta}}$,  $\mathbf{\lambda_{\varphi}}$, $\mathbf{\lambda_{\theta\varphi}}$ parameters have been evaluated in the helicity reference frame in the rapidity interval $-4 < y < -2.5$. The $\mathbf{\lambda_{\theta}}$ parameter has been found to be consistent with unity, and the other two parameters are found to be consistent with zero, indicating that coherently produced \jpsi mesons are transversely polarised as required for $s$-channel helicity conservation. The measured polarisation parameters are found to be compatible with those reported by the H1 Collaboration for \jpsi photoproduction in electron--proton interactions at lower energies, and evaluated in the helicity frame and using the spin density matrix formalism. This analysis represents the first experimental evidence that coherent \jpsi in ultra-peripheral \PbPb collisions at high energies is consistent with the photoproduction mechanism and the $s$-channel helicity conservation hypothesis.

%%%%%%%%%%%%%%%%%%%%%%%%%%%%%%%%
% end main text
%%%%%%%%%%%%%%%%%%%%%%%%%%%%%%%%

%%%%% acknowledgements - handled by EB chairs
\newenvironment{acknowledgement}{\relax}{\relax}
\begin{acknowledgement}
\section*{Acknowledgements}
% add specific acknowledgements here
% ...but please don't remove the line below: funding agencies
% will be acknowledged with a custom tex file handled by EB chairs after Collab Round 2
% Version: 2023-01-11

The ALICE Collaboration would like to thank all its engineers and technicians for their invaluable contributions to the construction of the experiment and the CERN accelerator teams for the outstanding performance of the LHC complex.
The ALICE Collaboration gratefully acknowledges the resources and support provided by all Grid centres and the Worldwide LHC Computing Grid (WLCG) collaboration.
The ALICE Collaboration acknowledges the following funding agencies for their support in building and running the ALICE detector:
A. I. Alikhanyan National Science Laboratory (Yerevan Physics Institute) Foundation (ANSL), State Committee of Science and World Federation of Scientists (WFS), Armenia;
Austrian Academy of Sciences, Austrian Science Fund (FWF): [M 2467-N36] and Nationalstiftung f\"{u}r Forschung, Technologie und Entwicklung, Austria;
Ministry of Communications and High Technologies, National Nuclear Research Center, Azerbaijan;
Conselho Nacional de Desenvolvimento Cient\'{\i}fico e Tecnol\'{o}gico (CNPq), Financiadora de Estudos e Projetos (Finep), Funda\c{c}\~{a}o de Amparo \`{a} Pesquisa do Estado de S\~{a}o Paulo (FAPESP) and Universidade Federal do Rio Grande do Sul (UFRGS), Brazil;
Bulgarian Ministry of Education and Science, within the National Roadmap for Research Infrastructures 2020-2027 (object CERN), Bulgaria;
Ministry of Education of China (MOEC) , Ministry of Science \& Technology of China (MSTC) and National Natural Science Foundation of China (NSFC), China;
Ministry of Science and Education and Croatian Science Foundation, Croatia;
Centro de Aplicaciones Tecnol\'{o}gicas y Desarrollo Nuclear (CEADEN), Cubaenerg\'{\i}a, Cuba;
Ministry of Education, Youth and Sports of the Czech Republic, Czech Republic;
The Danish Council for Independent Research | Natural Sciences, the VILLUM FONDEN and Danish National Research Foundation (DNRF), Denmark;
Helsinki Institute of Physics (HIP), Finland;
Commissariat \`{a} l'Energie Atomique (CEA) and Institut National de Physique Nucl\'{e}aire et de Physique des Particules (IN2P3) and Centre National de la Recherche Scientifique (CNRS), France;
Bundesministerium f\"{u}r Bildung und Forschung (BMBF) and GSI Helmholtzzentrum f\"{u}r Schwerionenforschung GmbH, Germany;
General Secretariat for Research and Technology, Ministry of Education, Research and Religions, Greece;
National Research, Development and Innovation Office, Hungary;
Department of Atomic Energy Government of India (DAE), Department of Science and Technology, Government of India (DST), University Grants Commission, Government of India (UGC) and Council of Scientific and Industrial Research (CSIR), India;
National Research and Innovation Agency - BRIN, Indonesia;
Istituto Nazionale di Fisica Nucleare (INFN), Italy;
Japanese Ministry of Education, Culture, Sports, Science and Technology (MEXT) and Japan Society for the Promotion of Science (JSPS) KAKENHI, Japan;
Consejo Nacional de Ciencia (CONACYT) y Tecnolog\'{i}a, through Fondo de Cooperaci\'{o}n Internacional en Ciencia y Tecnolog\'{i}a (FONCICYT) and Direcci\'{o}n General de Asuntos del Personal Academico (DGAPA), Mexico;
Nederlandse Organisatie voor Wetenschappelijk Onderzoek (NWO), Netherlands;
The Research Council of Norway, Norway;
Commission on Science and Technology for Sustainable Development in the South (COMSATS), Pakistan;
Pontificia Universidad Cat\'{o}lica del Per\'{u}, Peru;
Ministry of Education and Science, National Science Centre and WUT ID-UB, Poland;
Korea Institute of Science and Technology Information and National Research Foundation of Korea (NRF), Republic of Korea;
Ministry of Education and Scientific Research, Institute of Atomic Physics, Ministry of Research and Innovation and Institute of Atomic Physics and Universitatea Nationala de Stiinta si Tehnologie Politehnica Bucuresti, Romania;
Ministry of Education, Science, Research and Sport of the Slovak Republic, Slovakia;
National Research Foundation of South Africa, South Africa;
Swedish Research Council (VR) and Knut \& Alice Wallenberg Foundation (KAW), Sweden;
European Organization for Nuclear Research, Switzerland;
Suranaree University of Technology (SUT), National Science and Technology Development Agency (NSTDA) and National Science, Research and Innovation Fund (NSRF via PMU-B B05F650021), Thailand;
Turkish Energy, Nuclear and Mineral Research Agency (TENMAK), Turkey;
National Academy of  Sciences of Ukraine, Ukraine;
Science and Technology Facilities Council (STFC), United Kingdom;
National Science Foundation of the United States of America (NSF) and United States Department of Energy, Office of Nuclear Physics (DOE NP), United States of America.
In addition, individual groups or members have received support from:
Czech Science Foundation (grant no. 23-07499S), Czech Republic;
European Research Council, Strong 2020 - Horizon 2020, Marie Sk\l{}odowska Curie (grant nos. 950692, 824093, 896850), European Union;
ICSC - Centro Nazionale di Ricerca in High Performance Computing, Big Data and Quantum Computing, European Union - NextGenerationEU;
Academy of Finland (Center of Excellence in Quark Matter) (grant nos. 346327, 346328), Finland;
Programa de Apoyos para la Superaci\'{o}n del Personal Acad\'{e}mico, UNAM, Mexico.

\end{acknowledgement}

%%%%%%%% Bibliography
\bibliographystyle{utphys}   % Remember we use title in the biblio
\bibliography{bibliography}

\providecommand{\href}[2]{#2}\begingroup\raggedright\begin{thebibliography}{10}

\bibitem{Baltz:2007kq}
A.~J. Baltz, ``{The Physics of Ultraperipheral Collisions at the LHC}'',
  \href{https://doi.org/10.1016/j.physrep.2007.12.001}{{\em Phys. Rept.}
  {\bfseries 458} (2008) 1--171},
\href{https://arxiv.org/abs/0706.3356}{{\ttfamily arXiv:0706.3356 [nucl-ex]}}.
%%CITATION = ARXIV:0706.3356;%%.

\bibitem{Contreras:2015dqa}
J.~G. Contreras and J.~D. Tapia~Takaki, ``{Ultra-peripheral heavy-ion
  collisions at the LHC}'',
\href{https://doi.org/10.1142/S0217751X15420129}{{\em Int. J. Mod. Phys.}
  {\bfseries A30} (2015) 1542012}.
%%CITATION = IMPAE,A30,1542012;%%.

\bibitem{Klein:2020fmr}
S.~Klein and P.~Steinberg, ``{Photonuclear and Two-photon Interactions at
  High-Energy Nuclear Colliders}'',
  \href{https://doi.org/10.1146/annurev-nucl-030320-033923}{{\em Ann. Rev.
  Nucl. Part. Sci.} {\bfseries 70} (2020) 323--354},
  \href{https://arxiv.org/abs/2005.01872}{{\ttfamily arXiv:2005.01872
  [nucl-ex]}}.

\bibitem{Guzey:2016piu}
V.~Guzey, E.~Kryshen, and M.~Zhalov, ``{Coherent photoproduction of vector
  mesons in ultraperipheral heavy ion collisions: Update for run 2 at the CERN
  Large Hadron Collider}'',
  \href{https://doi.org/10.1103/PhysRevC.93.055206}{{\em Phys. Rev.} {\bfseries
  C93} (2016) 055206},
\href{https://arxiv.org/abs/1602.01456}{{\ttfamily arXiv:1602.01456
  [nucl-th]}}.
%%CITATION = ARXIV:1602.01456;%%.

\bibitem{Bendova:2020hbb}
D.~Bendova, J.~Cepila, J.~G. Contreras, and M.~Matas, ``{Photonuclear \jpsi
  production at the LHC: Proton-based versus nuclear dipole scattering
  amplitudes}'', \href{https://doi.org/10.1016/j.physletb.2021.136306}{{\em
  Phys. Lett. B} {\bfseries 817} (2021) 136306},
  \href{https://arxiv.org/abs/2006.12980}{{\ttfamily arXiv:2006.12980
  [hep-ph]}}.

\bibitem{Abelev:2012ba}
{\bfseries ALICE} Collaboration, B.~Abelev {\em et~al.}, ``{Coherent \jpsi
  photoproduction in ultra-peripheral \PbPb collisions at \twosevensixnn}'',
  \href{https://doi.org/10.1016/j.physletb.2012.11.059}{{\em Phys. Lett.}
  {\bfseries B718} (2013) 1273--1283},
\href{https://arxiv.org/abs/1209.3715}{{\ttfamily arXiv:1209.3715 [nucl-ex]}}.
%%CITATION = ARXIV:1209.3715;%%.

\bibitem{Abbas:2013oua}
{\bfseries ALICE} Collaboration, E.~Abbas {\em et~al.}, ``{Charmonium and
  e$^+$e$^-$ pair photoproduction at mid-rapidity in ultra-peripheral \PbPb
  collisions at \twosevensixnn}'',
  \href{https://doi.org/10.1140/epjc/s10052-013-2617-1}{{\em Eur. Phys. J.}
  {\bfseries C73} (2013) 2617},
\href{https://arxiv.org/abs/1305.1467}{{\ttfamily arXiv:1305.1467 [nucl-ex]}}.
%%CITATION = ARXIV:1305.1467;%%.

\bibitem{Khachatryan:2016qhq}
{\bfseries CMS} Collaboration, V.~Khachatryan {\em et~al.}, ``{Coherent \jpsi
  photoproduction in ultra-peripheral \PbPb collisions at $\sqrt {s_{\rm NN}}
  =$ 2.76 TeV with the CMS experiment}'',
  \href{https://doi.org/10.1016/j.physletb.2017.07.001}{{\em Phys. Lett.}
  {\bfseries B772} (2017) 489--511},
\href{https://arxiv.org/abs/1605.06966}{{\ttfamily arXiv:1605.06966
  [nucl-ex]}}.
%%CITATION = ARXIV:1605.06966;%%.

\bibitem{Tumasyan_2023}
{\bfseries CMS} Collaboration, A.~Tumasyan {\em et~al.}, ``Probing small
  {B}jorken-$x$ nuclear gluonic structure via coherent \jpsi photoproduction in
  ultraperipheral pb-pb collisions at \fivenn'',
  \href{https://doi.org/10.1103/physrevlett.131.262301}{{\em Physical Review
  Letters} {\bfseries 131} {262301}},
  \href{https://arxiv.org/abs/2303.16984}{{\ttfamily arXiv:2303.16984
  [hep-ex]}}. \url{http://dx.doi.org/10.1103/PhysRevLett.131.262301}.

\bibitem{LHCb:2021bfl}
{\bfseries LHCb} Collaboration, R.~Aaij {\em et~al.}, ``{Study of coherent
  \jpsi production in lead-lead collisions at \fiveExactly}'',
  \href{https://doi.org/10.1007/JHEP07(2022)117}{{\em JHEP} {\bfseries 07}
  (2022) 117}, \href{https://arxiv.org/abs/2107.03223}{{\ttfamily
  arXiv:2107.03223 [hep-ex]}}.

\bibitem{Guzey:2013xba}
V.~Guzey, E.~Kryshen, M.~Strikman, and M.~Zhalov, ``{Evidence for nuclear gluon
  shadowing from the ALICE measurements of \PbPb ultraperipheral exclusive
  \jpsi production}'',
  \href{https://doi.org/10.1016/j.physletb.2013.08.043}{{\em Phys. Lett.}
  {\bfseries B726} (2013) 290--295},
\href{https://arxiv.org/abs/1305.1724}{{\ttfamily arXiv:1305.1724 [hep-ph]}}.
%%CITATION = ARXIV:1305.1724;%%.

\bibitem{Gilman:1970vi}
F.~J. Gilman, J.~Pumplin, A.~Schwimmer, and L.~Stodolsky, ``{Helicity
  Conservation in Diffraction Scattering}'',
  \href{https://doi.org/10.1016/0370-2693(70)90203-0}{{\em Phys. Lett. B}
  {\bfseries 31} (1970) 387--390}.

\bibitem{Schilling:1973ag}
K.~Schilling and G.~Wolf, ``{How to analyze vector meson production in
  inelastic lepton scattering}'',
  \href{https://doi.org/10.1016/0550-3213(73)90371-4}{{\em Nucl. Phys. B}
  {\bfseries 61} (1973) 381--413}.

\bibitem{Ballam:1972eq}
J.~Ballam {\em et~al.}, ``{Vector Meson Production by Polarized Photons at 2.8
  GeV, 4.7 GeV, and 9.3 GeV}'',
  \href{https://doi.org/10.1103/PhysRevD.7.3150}{{\em Phys. Rev. D} {\bfseries
  7} (1973) 3150}.

\bibitem{OMEGAPhoton:1991exy}
{\bfseries OMEGA Photon} Collaboration, R.~J. Apsimon {\em et~al.},
  ``{Comparison of photon and hadron induced production of \rhozero mesons in
  the energy range of 65 GeV to 175 GeV}'',
  \href{https://doi.org/10.1007/BF01559734}{{\em Z. Phys. C} {\bfseries 53}
  (1992) 581--594}.

\bibitem{CLAS:2008rpm}
{\bfseries CLAS} Collaboration, S.~A. Morrow {\em et~al.}, ``{Exclusive
  \rhozero electroproduction on the proton at CLAS}'',
  \href{https://doi.org/10.1140/epja/i2008-10683-5}{{\em Eur. Phys. J. A}
  {\bfseries 39} (2009) 5--31},
  \href{https://arxiv.org/abs/0807.3834}{{\ttfamily arXiv:0807.3834 [hep-ex]}}.

\bibitem{Vernarsky:2014vaa}
{\bfseries CLAS} Collaboration, B.~Vernarsky, ``{First measurements of the
  $\rho^3$ spin density matrix elements in $\gamma {\rm p} \to {\rm p}\omega$
  using CLAS at JLAB}'', \href{https://doi.org/10.1142/S2010194514600635}{{\em
  Int. J. Mod. Phys. Conf. Ser.} {\bfseries 26} (2014) 1460063}.

\bibitem{Dey:2014tfa}
{\bfseries CLAS} Collaboration, B.~Dey {\em et~al.}, ``{Data analysis
  techniques, differential cross sections, and spin density matrix elements for
  the reaction $\gamma {\rm p} \rightarrow \phi {\rm p}$}'',
  \href{https://doi.org/10.1103/PhysRevC.89.055208}{{\em Phys. Rev. C}
  {\bfseries 89} (2014) 055208},
  \href{https://arxiv.org/abs/1403.2110}{{\ttfamily arXiv:1403.2110
  [nucl-ex]}}. [Addendum: Phys.Rev.C 90, 019901 (2014)].

\bibitem{Aktas:2005xu}
{\bfseries H1} Collaboration, A.~Aktas {\em et~al.}, ``{Elastic \jpsi
  production at HERA}'', \href{https://doi.org/10.1140/epjc/s2006-02519-5}{{\em
  Eur. Phys. J. C} {\bfseries 46} (2006) 585--603},
  \href{https://arxiv.org/abs/hep-ex/0510016}{{\ttfamily
  arXiv:hep-ex/0510016}}.

\bibitem{Chekanov:2004mw}
{\bfseries ZEUS} Collaboration, S.~Chekanov {\em et~al.}, ``{Exclusive
  electroproduction of \jpsi mesons at HERA}'',
  \href{https://doi.org/10.1016/j.nuclphysb.2004.06.034}{{\em Nucl. Phys. B}
  {\bfseries 695} (2004) 3--37},
  \href{https://arxiv.org/abs/hep-ex/0404008}{{\ttfamily
  arXiv:hep-ex/0404008}}.

\bibitem{ZEUS:1998fem}
{\bfseries ZEUS} Collaboration, J.~Breitweg {\em et~al.}, ``{Exclusive
  electroproduction of \rhozero and \jpsi mesons at HERA}'',
  \href{https://doi.org/10.1007/s100529901051}{{\em Eur. Phys. J. C} {\bfseries
  6} (1999) 603--627}, \href{https://arxiv.org/abs/hep-ex/9808020}{{\ttfamily
  arXiv:hep-ex/9808020}}.

\bibitem{Abelev_2008}
{\bfseries STAR} Collaboration, B.~I. Abelev {\em et~al.}, ``{\rhozero
  photoproduction in ultraperipheral relativistic heavy ion collisions at
  \twohundrednn}'', \href{https://doi.org/10.1103/PhysRevC.77.034910}{{\em
  Phys. Rev. C} {\bfseries 77} (2008) 034910},
  \href{https://arxiv.org/abs/0712.3320}{{\ttfamily arXiv:0712.3320
  [nucl-ex]}}.

\bibitem{Aamodt:2008zz}
{\bfseries ALICE} Collaboration, K.~Aamodt {\em et~al.}, ``{The ALICE
  experiment at the CERN LHC}'',
\href{https://doi.org/10.1088/1748-0221/3/08/S08002}{{\em JINST} {\bfseries 3}
  (2008) S08002}.
%%CITATION = JINST,3,S08002;%%.

\bibitem{Abelev:2014ffa}
{\bfseries ALICE} Collaboration, B.~Abelev {\em et~al.}, ``{Performance of the
  ALICE Experiment at the CERN LHC}'',
  \href{https://doi.org/10.1142/S0217751X14300440}{{\em Int. J. Mod. Phys.}
  {\bfseries A29} (2014) 1430044},
\href{https://arxiv.org/abs/1402.4476}{{\ttfamily arXiv:1402.4476 [nucl-ex]}}.
%%CITATION = ARXIV:1402.4476;%%.

\bibitem{Aamodt:2011gj}
{\bfseries ALICE} Collaboration, K.~Aamodt {\em et~al.}, ``{Rapidity and
  transverse momentum dependence of inclusive J$/\psi$ production in \pp
  collisions at $\sqrt{s} = 7$ TeV}'',
  \href{https://doi.org/10.1016/j.physletb.2011.09.054,
  10.1016/j.physletb.2012.10.060}{{\em Phys. Lett.} {\bfseries B704} (2011)
  442--455}, \href{https://arxiv.org/abs/1105.0380}{{\ttfamily arXiv:1105.0380
  [hep-ex]}}.
[Erratum: Phys. Lett.B718,692(2012)].
%%CITATION = ARXIV:1105.0380;%%.

\bibitem{Broz:2020ejr}
M.~Broz {\em et~al.}, ``{Performance of ALICE AD modules in the CERN PS test
  beam}'', \href{https://doi.org/10.1088/1748-0221/16/01/P01017}{{\em JINST}
  {\bfseries 16} (2021) P01017},
  \href{https://arxiv.org/abs/2006.14982}{{\ttfamily arXiv:2006.14982
  [physics.ins-det]}}.

\bibitem{N.Cartiglia:2015gve}
{\bfseries LHC Forward Physics Working Group} Collaboration, K.~Akiba {\em
  et~al.}, ``{LHC Forward Physics}'',
  \href{https://doi.org/10.1088/0954-3899/43/11/110201}{{\em J. Phys.}
  {\bfseries G43} (2016) 110201},
\href{https://arxiv.org/abs/1611.05079}{{\ttfamily arXiv:1611.05079 [hep-ph]}}.
%%CITATION = ARXIV:1611.05079;%%.

\bibitem{Acharya:2019vlb}
{\bfseries ALICE} Collaboration, S.~Acharya {\em et~al.}, ``{Coherent \jpsi
  photoproduction at forward rapidity in ultra-peripheral \PbPb collisions at
  \fivenn}'', \href{https://doi.org/10.1016/j.physletb.2019.134926}{{\em Phys.
  Lett. B} {\bfseries 798} (2019) 134926},
  \href{https://arxiv.org/abs/1904.06272}{{\ttfamily arXiv:1904.06272
  [nucl-ex]}}.

\bibitem{Bossu:2012jt}
{\bfseries ALICE} Collaboration, F.~Bossu, M.~Gagliardi, and M.~Marchisone,
  ``{Performance of the RPC-based ALICE muon trigger system at the LHC}'',
  \href{https://doi.org/10.1088/1748-0221/7/12/T12002}{{\em PoS} {\bfseries
  RPC2012} (2012) 059}, \href{https://arxiv.org/abs/1211.1948}{{\ttfamily
  arXiv:1211.1948 [physics.ins-det]}}.

\bibitem{ALICE:2012yye}
{\bfseries ALICE} Collaboration, B.~Abelev {\em et~al.}, ``{Coherent \jpsi
  photoproduction in ultra-peripheral \PbPb collisions at \twosevensixnn}'',
  \href{https://doi.org/10.1016/j.physletb.2012.11.059}{{\em Phys. Lett. B}
  {\bfseries 718} (2013) 1273--1283},
  \href{https://arxiv.org/abs/1209.3715}{{\ttfamily arXiv:1209.3715
  [nucl-ex]}}.

\bibitem{Klein:2016yzr}
S.~R. Klein, J.~Nystrand, J.~Seger, Y.~Gorbunov, and J.~Butterworth,
  ``{\starlight: A \montecarlo simulation program for ultra-peripheral
  collisions of relativistic ions}'',
  \href{https://doi.org/10.1016/j.cpc.2016.10.016}{{\em Comput. Phys. Commun.}
  {\bfseries 212} (2017) 258--268},
  \href{https://arxiv.org/abs/1607.03838}{{\ttfamily arXiv:1607.03838
  [hep-ph]}}.

\bibitem{Brun:1082634}
R.~Brun, F.~Bruyant, F.~Carminati, S.~Giani, M.~Maire, A.~McPherson,
  G.~Patrick, and L.~Urban, \href{https://doi.org/10.17181/CERN.MUHF.DMJ1}{{\em
  {GEANT: Detector Description and Simulation Tool; Oct 1994}}}.
\newblock CERN Program Library. CERN, Geneva, 1993.
\newblock \url{http://cds.cern.ch/record/1082634}.
\newblock Long Writeup W5013.

\bibitem{Faccioli:2010kd}
P.~Faccioli, C.~Louren{\c c}o, J.~Seixas, and H.~K. W{\"o}hri, ``{Towards the
  experimental clarification of quarkonium polarization}'',
  \href{https://doi.org/10.1140/epjc/s10052-010-1420-5}{{\em Eur. Phys. J. C}
  {\bfseries 69} (2010) 657--673},
  \href{https://arxiv.org/abs/1006.2738}{{\ttfamily arXiv:1006.2738 [hep-ph]}}.

\bibitem{Abelev:2011md}
{\bfseries ALICE} Collaboration, B.~Abelev {\em et~al.}, ``{\jpsi polarization
  in \pp collisions at \seven}'',
  \href{https://doi.org/10.1103/PhysRevLett.108.082001}{{\em Phys. Rev. Lett.}
  {\bfseries 108} (2012) 082001},
  \href{https://arxiv.org/abs/1111.1630}{{\ttfamily arXiv:1111.1630 [hep-ex]}}.

\bibitem{Acharya:2018uww}
{\bfseries ALICE} Collaboration, S.~Acharya {\em et~al.}, ``{Measurement of the
  inclusive \jpsi polarization at forward rapidity in \pp collisions at
  \eight}'', \href{https://doi.org/10.1140/epjc/s10052-018-6027-2}{{\em Eur.
  Phys. J. C} {\bfseries 78} (2018) 562},
  \href{https://arxiv.org/abs/1805.04374}{{\ttfamily arXiv:1805.04374
  [hep-ex]}}.

\bibitem{Acharya:2020xko}
{\bfseries ALICE} Collaboration, S.~Acharya {\em et~al.}, ``{First measurement
  of quarkonium polarization in nuclear collisions at the LHC}'',
  \href{https://doi.org/10.1016/j.physletb.2021.136146}{{\em Phys. Lett. B}
  {\bfseries 815} (2021) 136146},
  \href{https://arxiv.org/abs/2005.11128}{{\ttfamily arXiv:2005.11128
  [nucl-ex]}}.

\bibitem{ALICE:2022sli}
{\bfseries ALICE} Collaboration, S.~Acharya {\em et~al.}, ``{Measurement of the
  \jpsi polarization with respect to the event plane in \PbPb collisions at the
  LHC}'', \href{https://doi.org/10.1103/PhysRevLett.131.042303}{{\em Phys. Rev.
  Lett.} {\bfseries 131} (2023) 042303},
  \href{https://arxiv.org/abs/2204.10171}{{\ttfamily arXiv:2204.10171
  [nucl-ex]}}.

\bibitem{Acharya:2020oym}
{\bfseries PHENIX} Collaboration, U.~Acharya {\em et~al.}, ``{Polarization and
  cross section of midrapidity \jpsi production in p$+$p collisions at $\sqrt
  {s}$ = 510 GeV}'', \href{https://doi.org/10.1103/PhysRevD.102.072008}{{\em
  Phys. Rev. D} {\bfseries 102} (2020) 072008},
  \href{https://arxiv.org/abs/2005.14273}{{\ttfamily arXiv:2005.14273
  [hep-ex]}}.

\bibitem{Faccioli:2010ji}
P.~Faccioli, C.~Louren{\c c}o, and J.~Seixas, ``{A New approach to quarkonium
  polarization studies}'',
  \href{https://doi.org/10.1103/PhysRevD.81.111502}{{\em Phys. Rev.} {\bfseries
  D81} (2010) 111502},
\href{https://arxiv.org/abs/1005.2855}{{\ttfamily arXiv:1005.2855 [hep-ph]}}.
%%CITATION = ARXIV:1005.2855;%%.

\bibitem{Gaiser:1982yw}
J.~E. Gaiser, {\em {Charmonium Spectroscopy From Radiative Decays of the \jpsi
  and \psip}}.
\newblock PhD thesis, SLAC, 1982.
\newblock
\url{http://inspirehep.net/record/183554/files/slac-r-255.pdf}.
\newblock
%%CITATION = SLAC-0255;%%.

\bibitem{Tanabashi:2018oca}
{\bfseries Particle Data Group} Collaboration, M.~Tanabashi {\em et~al.},
  ``{Review of Particle Physics}'',
\href{https://doi.org/10.1103/PhysRevD.98.030001}{{\em Phys. Rev.} {\bfseries
  D98} (2018) 030001}.
%%CITATION = PHRVA,D98,030001;%%.

\bibitem{Alexa:2013xxa}
{\bfseries H1} Collaboration, C.~Alexa {\em et~al.}, ``{Elastic and
  Proton-Dissociative Photoproduction of \jpsi Mesons at HERA}'',
  \href{https://doi.org/10.1140/epjc/s10052-013-2466-y}{{\em Eur. Phys. J.}
  {\bfseries C73} (2013) 2466},
\href{https://arxiv.org/abs/1304.5162}{{\ttfamily arXiv:1304.5162 [hep-ex]}}.
%%CITATION = ARXIV:1304.5162;%%.

\bibitem{DAgostini:1994fjx}
G.~D'Agostini, ``{A Multidimensional unfolding method based on Bayes'
  theorem}'', \href{https://doi.org/10.1016/0168-9002(95)00274-X}{{\em Nucl.
  Instrum. Meth. A} {\bfseries 362} (1995) 487--498}.

\bibitem{Adye:2011gm}
T.~Adye, \href{https://doi.org/10.5170/CERN-2011-006.313}{``{Unfolding
  algorithms and tests using RooUnfold}'',} in {\em {PHYSTAT 2011}},
  pp.~313--318.
\newblock CERN, Geneva, 2011.
\newblock \href{https://arxiv.org/abs/1105.1160}{{\ttfamily arXiv:1105.1160
  [physics.data-an]}}.

\bibitem{Schmitt:2016orm}
S.~Schmitt, ``{Data Unfolding Methods in High Energy Physics}'',
  \href{https://doi.org/10.1051/epjconf/201713711008}{{\em EPJ Web Conf.}
  {\bfseries 137} (2017) 11008},
  \href{https://arxiv.org/abs/1611.01927}{{\ttfamily arXiv:1611.01927
  [physics.data-an]}}.

\bibitem{Drijard:318006}
D.~Drijard, W.~T. Eadie, F.~E. James, M.~Roos, and B.~Sadoulet, ``{Probability
  and statistics applied to high-energy physics}'', {\em CERN Academic Training
  Lectures} {\bfseries 39, {\rm vol. 7}} (1968) 1.
  \url{https://cds.cern.ch/record/318006}. CERN, Geneva, 1968 - 1969.

\bibitem{ParticleDataGroup:2024cfk}
{\bfseries Particle Data Group} Collaboration, S.~Navas {\em et~al.}, ``{Review
  of particle physics}'',
  \href{https://doi.org/10.1103/PhysRevD.110.030001}{{\em Phys. Rev. D}
  {\bfseries 110} (2024) 030001, {\rm Section 40.4.2.3}}.

\bibitem{Martens:2017cvj}
J.~C. Martens, J.~P. Ralston, and J.~D. Tapia~Takaki, ``{Quantum tomography for
  collider physics: Illustrations with lepton pair production}'',
  \href{https://doi.org/10.1140/epjc/s10052-017-5455-8}{{\em Eur. Phys. J. C}
  {\bfseries 78} (2018) 5}, \href{https://arxiv.org/abs/1707.01638}{{\ttfamily
  arXiv:1707.01638 [hep-ph]}}.

\bibitem{Gavrilova:2019jea}
M.~Gavrilova and O.~Teryaev, ``{Rotation-invariant observables as Density
  Matrix invariants}'', \href{https://doi.org/10.1103/PhysRevD.99.076013}{{\em
  Phys. Rev. D} {\bfseries 99} (2019) 076013},
  \href{https://arxiv.org/abs/1901.04018}{{\ttfamily arXiv:1901.04018
  [hep-ph]}}.

\end{thebibliography}\endgroup
%\input {bibliography.tex}

%%%%%%%%%%%%%%%%%%%%%%%%%%%%%%%%
% Appendices: yours (if any) + authorlist
%%%%%%%%%%%%%%%%%%%%%%%%%%%%%%%%
\newpage
\appendix

%
%\input{} % put your appendices here (if any)
%

%%%%% Authorlist - please do not touch: handled by EB chairs
\section{The ALICE Collaboration}
\label{app:collab}
% ALICE Collaboration author list for 2023-01-11
\begin{flushleft} 
\small

S.~Acharya\,\orcidlink{0000-0002-9213-5329}\,$^{\rm 126}$, 
D.~Adamov\'{a}\,\orcidlink{0000-0002-0504-7428}\,$^{\rm 86}$, 
A.~Adler$^{\rm 70}$, 
G.~Aglieri Rinella\,\orcidlink{0000-0002-9611-3696}\,$^{\rm 32}$, 
M.~Agnello\,\orcidlink{0000-0002-0760-5075}\,$^{\rm 29}$, 
N.~Agrawal\,\orcidlink{0000-0003-0348-9836}\,$^{\rm 51}$, 
Z.~Ahammed\,\orcidlink{0000-0001-5241-7412}\,$^{\rm 134}$, 
S.~Ahmad\,\orcidlink{0000-0003-0497-5705}\,$^{\rm 15}$, 
S.U.~Ahn\,\orcidlink{0000-0001-8847-489X}\,$^{\rm 71}$, 
I.~Ahuja\,\orcidlink{0000-0002-4417-1392}\,$^{\rm 37}$
% ...

\section*{Affiliation Notes}

$^{\rm I}$ Deceased\\
$^{\rm II}$ Also at: Max-Planck-Institut fur Physik, Munich, Germany\\
$^{\rm III}$ Also at: Italian National Agency for New Technologies, Energy and Sustainable Economic Development (ENEA), Bologna, Italy\\
$^{\rm IV}$ Also at: Dipartimento DET del Politecnico di Torino, Turin, Italy\\
$^{\rm V}$ Also at: Yildiz Technical University, Istanbul, T\"{u}rkiye\\
$^{\rm VI}$ Also at: Department of Applied Physics, Aligarh Muslim University, Aligarh, India\\
$^{\rm VII}$ Also at: Institute of Theoretical Physics, University of Wroclaw, Poland\\
$^{\rm VIII}$ Also at: An institution covered by a cooperation agreement with CERN\\

\section*{Collaboration Institutes}

$^{1}$ A.I. Alikhanyan National Science Laboratory (Yerevan Physics Institute) Foundation, Yerevan, Armenia\\
$^{2}$ AGH University of Krakow, Cracow, Poland\\
$^{3}$ Bogolyubov Institute for Theoretical Physics, National Academy of Sciences of Ukraine, Kiev, Ukraine\\
$^{4}$ Bose Institute, Department of Physics  and Centre for Astroparticle Physics and Space Science (CAPSS), Kolkata, India\\
$^{5}$ California Polytechnic State University, San Luis Obispo, California, United States\\
$^{6}$ Central China Normal University, Wuhan, China\\
$^{7}$ Centro de Aplicaciones Tecnol\'{o}gicas y Desarrollo Nuclear (CEADEN), Havana, Cuba\\
$^{8}$ Centro de Investigaci\'{o}n y de Estudios Avanzados (CINVESTAV), Mexico City and M\'{e}rida, Mexico\\
$^{9}$ Chicago State University, Chicago, Illinois, United States\\
$^{10}$ China Institute of Atomic Energy, Beijing, China\\
$^{11}$ Chungbuk National University, Cheongju, Republic of Korea\\
$^{12}$ Comenius University Bratislava, Faculty of Mathematics, Physics and Informatics, Bratislava, Slovak Republic\\
$^{13}$ COMSATS University Islamabad, Islamabad, Pakistan\\
$^{14}$ Creighton University, Omaha, Nebraska, United States\\
$^{15}$ Department of Physics, Aligarh Muslim University, Aligarh, India\\
$^{16}$ Department of Physics, Pusan National University, Pusan, Republic of Korea\\
$^{17}$ Department of Physics, Sejong University, Seoul, Republic of Korea\\
$^{18}$ Department of Physics, University of California, Berkeley, California, United States\\
$^{19}$ Department of Physics, University of Oslo, Oslo, Norway\\
$^{20}$ Department of Physics and Technology, University of Bergen, Bergen, Norway\\
$^{21}$ Dipartimento di Fisica, Universit\`{a} di Pavia, Pavia, Italy\\
$^{22}$ Dipartimento di Fisica dell'Universit\`{a} and Sezione INFN, Cagliari, Italy\\
$^{23}$ Dipartimento di Fisica dell'Universit\`{a} and Sezione INFN, Trieste, Italy\\
$^{24}$ Dipartimento di Fisica dell'Universit\`{a} and Sezione INFN, Turin, Italy\\
$^{25}$ Dipartimento di Fisica e Astronomia dell'Universit\`{a} and Sezione INFN, Bologna, Italy\\
$^{26}$ Dipartimento di Fisica e Astronomia dell'Universit\`{a} and Sezione INFN, Catania, Italy\\
$^{27}$ Dipartimento di Fisica e Astronomia dell'Universit\`{a} and Sezione INFN, Padova, Italy\\
$^{28}$ Dipartimento di Fisica `E.R.~Caianiello' dell'Universit\`{a} and Gruppo Collegato INFN, Salerno, Italy\\
$^{29}$ Dipartimento DISAT del Politecnico and Sezione INFN, Turin, Italy\\
$^{30}$ Dipartimento di Scienze MIFT, Universit\`{a} di Messina, Messina, Italy\\
$^{31}$ Dipartimento Interateneo di Fisica `M.~Merlin' and Sezione INFN, Bari, Italy\\
$^{32}$ European Organization for Nuclear Research (CERN), Geneva, Switzerland\\
$^{33}$ Faculty of Electrical Engineering, Mechanical Engineering and Naval Architecture, University of Split, Split, Croatia\\
$^{34}$ Faculty of Engineering and Science, Western Norway University of Applied Sciences, Bergen, Norway\\
$^{35}$ Faculty of Nuclear Sciences and Physical Engineering, Czech Technical University in Prague, Prague, Czech Republic\\
$^{36}$ Faculty of Physics, Sofia University, Sofia, Bulgaria\\
$^{37}$ Faculty of Science, P.J.~\v{S}af\'{a}rik University, Ko\v{s}ice, Slovak Republic\\
$^{38}$ Frankfurt Institute for Advanced Studies, Johann Wolfgang Goethe-Universit\"{a}t Frankfurt, Frankfurt, Germany\\
$^{39}$ Fudan University, Shanghai, China\\
$^{40}$ Gangneung-Wonju National University, Gangneung, Republic of Korea\\
$^{41}$ Gauhati University, Department of Physics, Guwahati, India\\
$^{42}$ Helmholtz-Institut f\"{u}r Strahlen- und Kernphysik, Rheinische Friedrich-Wilhelms-Universit\"{a}t Bonn, Bonn, Germany\\
$^{43}$ Helsinki Institute of Physics (HIP), Helsinki, Finland\\
$^{44}$ High Energy Physics Group,  Universidad Aut\'{o}noma de Puebla, Puebla, Mexico\\
$^{45}$ Horia Hulubei National Institute of Physics and Nuclear Engineering, Bucharest, Romania\\
$^{46}$ HUN-REN Wigner Research Centre for Physics, Budapest, Hungary\\
$^{47}$ Indian Institute of Technology Bombay (IIT), Mumbai, India\\
$^{48}$ Indian Institute of Technology Indore, Indore, India\\
$^{49}$ INFN, Laboratori Nazionali di Frascati, Frascati, Italy\\
$^{50}$ INFN, Sezione di Bari, Bari, Italy\\
$^{51}$ INFN, Sezione di Bologna, Bologna, Italy\\
$^{52}$ INFN, Sezione di Cagliari, Cagliari, Italy\\
$^{53}$ INFN, Sezione di Catania, Catania, Italy\\
$^{54}$ INFN, Sezione di Padova, Padova, Italy\\
$^{55}$ INFN, Sezione di Pavia, Pavia, Italy\\
$^{56}$ INFN, Sezione di Torino, Turin, Italy\\
$^{57}$ INFN, Sezione di Trieste, Trieste, Italy\\
$^{58}$ Inha University, Incheon, Republic of Korea\\
$^{59}$ Institute for Gravitational and Subatomic Physics (GRASP), Utrecht University/Nikhef, Utrecht, Netherlands\\
$^{60}$ Institute of Experimental Physics, Slovak Academy of Sciences, Ko\v{s}ice, Slovak Republic\\
$^{61}$ Institute of Physics, Homi Bhabha National Institute, Bhubaneswar, India\\
$^{62}$ Institute of Physics of the Czech Academy of Sciences, Prague, Czech Republic\\
$^{63}$ Institute of Space Science (ISS), Bucharest, Romania\\
$^{64}$ Institut f\"{u}r Kernphysik, Johann Wolfgang Goethe-Universit\"{a}t Frankfurt, Frankfurt, Germany\\
$^{65}$ Instituto de Ciencias Nucleares, Universidad Nacional Aut\'{o}noma de M\'{e}xico, Mexico City, Mexico\\
$^{66}$ Instituto de F\'{i}sica, Universidade Federal do Rio Grande do Sul (UFRGS), Porto Alegre, Brazil\\
$^{67}$ Instituto de F\'{\i}sica, Universidad Nacional Aut\'{o}noma de M\'{e}xico, Mexico City, Mexico\\
$^{68}$ iThemba LABS, National Research Foundation, Somerset West, South Africa\\
$^{69}$ Jeonbuk National University, Jeonju, Republic of Korea\\
$^{70}$ Johann-Wolfgang-Goethe Universit\"{a}t Frankfurt Institut f\"{u}r Informatik, Fachbereich Informatik und Mathematik, Frankfurt, Germany\\
$^{71}$ Korea Institute of Science and Technology Information, Daejeon, Republic of Korea\\
$^{72}$ KTO Karatay University, Konya, Turkey\\
$^{73}$ Laboratoire de Physique Subatomique et de Cosmologie, Universit\'{e} Grenoble-Alpes, CNRS-IN2P3, Grenoble, France\\
$^{74}$ Lawrence Berkeley National Laboratory, Berkeley, California, United States\\
$^{75}$ Lund University Department of Physics, Division of Particle Physics, Lund, Sweden\\
$^{76}$ Nagasaki Institute of Applied Science, Nagasaki, Japan\\
$^{77}$ Nara Women{'}s University (NWU), Nara, Japan\\
$^{78}$ National and Kapodistrian University of Athens, School of Science, Department of Physics , Athens, Greece\\
$^{79}$ National Centre for Nuclear Research, Warsaw, Poland\\
$^{80}$ National Institute of Science Education and Research, Homi Bhabha National Institute, Jatni, India\\
$^{81}$ National Nuclear Research Center, Baku, Azerbaijan\\
$^{82}$ National Research and Innovation Agency - BRIN, Jakarta, Indonesia\\
$^{83}$ Niels Bohr Institute, University of Copenhagen, Copenhagen, Denmark\\
$^{84}$ Nikhef, National institute for subatomic physics, Amsterdam, Netherlands\\
$^{85}$ Nuclear Physics Group, STFC Daresbury Laboratory, Daresbury, United Kingdom\\
$^{86}$ Nuclear Physics Institute of the Czech Academy of Sciences, Husinec-\v{R}e\v{z}, Czech Republic\\
$^{87}$ Oak Ridge National Laboratory, Oak Ridge, Tennessee, United States\\
$^{88}$ Ohio State University, Columbus, Ohio, United States\\
$^{89}$ Physics department, Faculty of science, University of Zagreb, Zagreb, Croatia\\
$^{90}$ Physics Department, Panjab University, Chandigarh, India\\
$^{91}$ Physics Department, University of Jammu, Jammu, India\\
$^{92}$ Physics Program and International Institute for Sustainability with Knotted Chiral Meta Matter (SKCM2), Hiroshima University, Hiroshima, Japan\\
$^{93}$ Physikalisches Institut, Eberhard-Karls-Universit\"{a}t T\"{u}bingen, T\"{u}bingen, Germany\\
$^{94}$ Physikalisches Institut, Ruprecht-Karls-Universit\"{a}t Heidelberg, Heidelberg, Germany\\
$^{95}$ Physik Department, Technische Universit\"{a}t M\"{u}nchen, Munich, Germany\\
$^{96}$ Politecnico di Bari and Sezione INFN, Bari, Italy\\
$^{97}$ Research Division and ExtreMe Matter Institute EMMI, GSI Helmholtzzentrum f\"ur Schwerionenforschung GmbH, Darmstadt, Germany\\
$^{98}$ Saga University, Saga, Japan\\
$^{99}$ Saha Institute of Nuclear Physics, Homi Bhabha National Institute, Kolkata, India\\
$^{100}$ School of Physics and Astronomy, University of Birmingham, Birmingham, United Kingdom\\
$^{101}$ Secci\'{o}n F\'{\i}sica, Departamento de Ciencias, Pontificia Universidad Cat\'{o}lica del Per\'{u}, Lima, Peru\\
$^{102}$ Stefan Meyer Institut f\"{u}r Subatomare Physik (SMI), Vienna, Austria\\
$^{103}$ SUBATECH, IMT Atlantique, Nantes Universit\'{e}, CNRS-IN2P3, Nantes, France\\
$^{104}$ Sungkyunkwan University, Suwon City, Republic of Korea\\
$^{105}$ Suranaree University of Technology, Nakhon Ratchasima, Thailand\\
$^{106}$ Technical University of Ko\v{s}ice, Ko\v{s}ice, Slovak Republic\\
$^{107}$ The Henryk Niewodniczanski Institute of Nuclear Physics, Polish Academy of Sciences, Cracow, Poland\\
$^{108}$ The University of Texas at Austin, Austin, Texas, United States\\
$^{109}$ Universidad Aut\'{o}noma de Sinaloa, Culiac\'{a}n, Mexico\\
$^{110}$ Universidade de S\~{a}o Paulo (USP), S\~{a}o Paulo, Brazil\\
$^{111}$ Universidade Estadual de Campinas (UNICAMP), Campinas, Brazil\\
$^{112}$ Universidade Federal do ABC, Santo Andre, Brazil\\
$^{113}$ Universitatea Nationala de Stiinta si Tehnologie Politehnica Bucuresti, Bucharest, Romania\\
$^{114}$ University of Cape Town, Cape Town, South Africa\\
$^{115}$ University of Houston, Houston, Texas, United States\\
$^{116}$ University of Jyv\"{a}skyl\"{a}, Jyv\"{a}skyl\"{a}, Finland\\
$^{117}$ University of Kansas, Lawrence, Kansas, United States\\
$^{118}$ University of Liverpool, Liverpool, United Kingdom\\
$^{119}$ University of Science and Technology of China, Hefei, China\\
$^{120}$ University of South-Eastern Norway, Kongsberg, Norway\\
$^{121}$ University of Tennessee, Knoxville, Tennessee, United States\\
$^{122}$ University of the Witwatersrand, Johannesburg, South Africa\\
$^{123}$ University of Tokyo, Tokyo, Japan\\
$^{124}$ University of Tsukuba, Tsukuba, Japan\\
$^{125}$ Universit\"{a}t M\"{u}nster, Institut f\"{u}r Kernphysik, M\"{u}nster, Germany\\
$^{126}$ Universit\'{e} Clermont Auvergne, CNRS/IN2P3, LPC, Clermont-Ferrand, France\\
$^{127}$ Universit\'{e} de Lyon, CNRS/IN2P3, Institut de Physique des 2 Infinis de Lyon, Lyon, France\\
$^{128}$ Universit\'{e} de Strasbourg, CNRS, IPHC UMR 7178, F-67000 Strasbourg, France, Strasbourg, France\\
$^{129}$ Universit\'{e} Paris-Saclay, Centre d'Etudes de Saclay (CEA), IRFU, D\'{e}partment de Physique Nucl\'{e}aire (DPhN), Saclay, France\\
$^{130}$ Universit\'{e}  Paris-Saclay, CNRS/IN2P3, IJCLab, Orsay, France\\
$^{131}$ Universit\`{a} degli Studi di Foggia, Foggia, Italy\\
$^{132}$ Universit\`{a} del Piemonte Orientale, Vercelli, Italy\\
$^{133}$ Universit\`{a} di Brescia, Brescia, Italy\\
$^{134}$ Variable Energy Cyclotron Centre, Homi Bhabha National Institute, Kolkata, India\\
$^{135}$ Warsaw University of Technology, Warsaw, Poland\\
$^{136}$ Wayne State University, Detroit, Michigan, United States\\
$^{137}$ Yale University, New Haven, Connecticut, United States\\
$^{138}$ Yonsei University, Seoul, Republic of Korea\\
$^{139}$  Zentrum  f\"{u}r Technologie und Transfer (ZTT), Worms, Germany\\
$^{140}$ Affiliated with an institute covered by a cooperation agreement with CERN\\
$^{141}$ Affiliated with an international laboratory covered by a cooperation agreement with CERN.\\

\end{flushleft} 

\end{document}